\documentclass[aps,prd,twocolumn,a4paper,superscriptaddress,10pt,floatfix]{revtex4-2}
\usepackage{graphicx,multirow,latexsym,hyperref,amsmath,amsmath}
\usepackage[british]{babel}

\begin{document}
\newcommand{\be}{\begin{equation}}
\newcommand{\ee}{\end{equation}}
\newcommand{\bq}{\begin{eqnarray}}
\newcommand{\eq}{\end{eqnarray}}
\newcommand{\bsq}{\begin{subequations}}
\newcommand{\esq}{\end{subequations}}
\newcommand{\bc}{\begin{center}}
\newcommand{\ec}{\end{center}}

\title{Probing Unification Scenarios with Big Bang Nucleosynthesis}
\author{I. M. Dreyer}
\email{iunamdreyer@tecnico.ulisboa.pt}
\affiliation{CENTRA, Departamento de F\'{\i}sica, Instituto Superior T\'ecnico,
Universidade de Lisboa, Avenida Rovisco Pais 1, 1049-001 Lisboa, Portugal}
\affiliation{Centro de Astrof\'{\i}sica da Universidade do Porto, Rua das Estrelas, 4150-762 Porto, Portugal}
\author{C. J. A. P. Martins}
\email{Carlos.Martins@astro.up.pt}
\affiliation{Centro de Astrof\'{\i}sica da Universidade do Porto, Rua das Estrelas, 4150-762 Porto, Portugal}
\affiliation{Instituto de Astrof\'{\i}sica e Ci\^encias do Espa\c co, Universidade do Porto, Rua das Estrelas, 4150-762 Porto, Portugal}
\date{\today}

\begin{abstract}
We extend a recently developed Big Bang Nucleosynthesis (BBN) code, {\tt PRyMordial}, to constrain a broad class of Grand Unified Theories to which BBN is sensitive, since these lead to varying fundamental couplings. A previously developed self-consistent perturbative analysis of the effects of these variations has been implemented in {\tt PRyMordial}, leading to robust constraints of the value of the fine-structure constant, $\alpha$, at the BBN epoch using current observations of Helium-4 and Deuterium abundances. We explored two different viable scenarios, relying on alternative assumptions on the gravitational sector: the variation of the gravitational coupling can be implemented by varying either particle masses, or Newton's gravitational constant. For the variation of masses, we obtained at $68\%$ confidence level a constraint on the relative variation of $\alpha$, between the BBN epoch and the present-day laboratory value, of $\Delta\alpha/\alpha=2\pm51$ ppm (parts per million), while for the variation of Newton's constant the analogous constraint is $\Delta\alpha/\alpha=2\pm22$ ppm. We also show that, given these constraints, these models do not provide a solution to the cosmological Lithium problem.
\end{abstract}
\maketitle
\section{\label{section1}Introduction}

Big Bang Nucleosynthesis (BBN) is an observational cornerstone of the Hot Big Bang model and a sensitive probe of physics beyond it \cite{Sarkar:1995dd,Steigman:2007xt,Iocco:2008va,primatHe4}. It is the epoch when protons and neutrons formed the first light atomic nuclei---Hydrogen (${}^1$H), Deuterium (${}^2$H, D), Helium-3 (${}^3$He), Helium-4 (${}^4$He) and Lithium-7 (${}^7$Li). Extrapolation of observational data from low metallicity regions of the Universe, together with current knowledge of thermodynamics and nuclear reaction rates allows BBN abundances to be predicted with good precision and accuracy. Although some analytic approximations can be made, a fully consistent analysis must be done numerically \cite{Kawano1,Kawano2,Pisanti,primatHe4}.

Grand Unified Theories (GUT) are a class of models that describe the unification of the three particle physics fundamental forces (strong, electromagnetic and weak). According to Quantum Field Theory, to achieve this unification, the coupling strengths $\alpha_i$ of the different forces have to run with energy, until they converge at some very high energy scale. They may also have additional spacetime dependencies, e.g. driven by dynamical scalar fields. To verify this kind of theory, an environment with extreme energies is needed. One such possibility relies on particle accelerators, but an alternative is provided by the early Universe. As a well-studied era with good simulations and increasingly precise and accurate data, especially for Deuterium, BBN becomes an ideal tool to study the effects of coupling variations and constrain GUT models. Moreover, BBN is fundamentally a competition between nuclear reaction rates, the neutron lifetime, and the Universe's expansion rate, implying that all four fundamental forces play a role in its outcome, and it is therefore well suited to probe these models, where multiple couplings are varying. Inter alia, we will see that the assumptions on the nuclear reaction rates---specifically, whether one use {\tt NACRE II} \cite{nacre} or {\tt PRIMAT} \cite{primatHe4} databases---has some impact on the results.

In this work we take the recently published open source Python code {\tt PRyMordial} \cite{Burns}, and extend it by self-consistently implementing coupling variations occurring in a broad GUT class into this code, following a previously developed perturbative analysis \cite{Martins1}. The perturbations are always done with respect to the standard values of cosmological and nuclear physical parameters, and the assumed observed values of primordial abundances are the ones recommended in the BBN review article in \cite{PDG}. Specifically, we assume that the fine-structure constant, $\alpha$, may have had a value at the BBN that does not coincide with the present-day laboratory one, $\alpha_0$, but this value was nevertheless constant during the relevant BBN period. This difference will be quantified through the relative variation, defined as $\Delta\alpha/\alpha=(\alpha_{BBN}-\alpha_0)/\alpha_0$. Other quantities also vary, and their relative variations are parametrically related to $\Delta\alpha/\alpha$, as described in the next section.

The plan of the rest of this work is as follows. Section \ref{section2} contains a brief overview of the class of GUT models under consideration and its perturbative treatment. Its implementation in the {\tt PRyMordial} code is then summarized in Sect. \ref{section3}, with some further details in Appendix \ref{app1} and additional ones available in \cite{Thesis}. Section \ref{section4} contains an initial comparison of these models with observations, aiming to identify the parameter ranges leading to possibly viable models, while a more thorough statistical analysis, and the resulting constraints on this class of models, are discussed in Sect. \ref{section5}. In Sect. \ref{section6} we briefly revisit the Lithium problem in the context of our results, Finally, Sect. \ref{section7} contains our conclusions.

\section{\label{section2}Perturbative analysis}

A general, self-contained perturbative approach to the effects of variations of nature’s fundamental constants in a broad class of GUTs has been introduced in \cite{Martins1,Martins2}, drawing on previous work by other groups \cite{Coc,Dent}. These are the main sources for the following expressions, with some additional details from  \cite{Luo,kappas,WhiteDwarfs}.

Our goal is to be able to constrain models which allow for simultaneous variations of several fundamental couplings. For this to be practical (with a manageable number of free parameters), one needs to relate the various changes to those of a particular dimensionless coupling, conveniently chosen to be the fine structure constant $\alpha_{EM}=e^2/4\pi\varepsilon_0\hbar c$ (henceforth denoted simply $\alpha$). Following \cite{Coc} we consider a broad class of GUTs where the weak scale is determined by dimensional transmutation, further assuming that the relative variation of all the Yukawa couplings is the same and that the variation of the couplings is driven by a dilaton-type scalar field \cite{Campbell}, which we do not need to explicitly specify, other than assuming that it is not dominating the cosmological dynamics at early times, and therefore its only direct cosmological impact is a possible change in Newton's gravitational constant (on which more below). 

Regarding variations of dimensionful quantities with units of mass (or energy), one starts with the gravitational fine structure constant $\alpha_g=G_N m_p^2/4\pi\varepsilon_0\hbar c$. (The inclusion of the proton mass in this definition is the canonical one; one could define analogous quantities with the neutron or electron mass, but free neutrons are unstable and electrons are much lighter than protons.) What can be directly measured, by observation, is the variation of this dimensionless $\alpha_g$, and if there is such a variation, it can be modeled by a variation in $G_N$ or by a variation of the particle masses. Crucially, $\alpha_g$ is proportional to the square of the ratio of the QCD mass scale and the Planck mass. Therefore, one can either stipulate the QCD scale and particle masses vary while the Planck mass is fixed (as was assumed in \cite{Coc}), or keep the QCD mass scale fixed and allow the Planck mass to vary (as was assumed in \cite{Dent}). In practice, the variation of a generic mass $m_x$, measured against a reference mass $m_r$, will be $\Delta m_x/m_x\equiv((m_x/m_r)_{BBN}-(m_x/m_r)_0)/(m_x/m_r)_0$.

Let's start with the choice that the QCD scale and particle masses vary while the Planck mass is fixed, i.e. setting $m_r=M_{Pl}$. The above assumptions suffice to self-consistently relate the relative variations of all the parameters impacting BBN, using only two dimensionless parameters in addition to $\alpha$: one of these ($R$) is related to the strong sector, while the other ($S$) is related to the electroweak sector. Specifically, these are defined as
\bq
\frac{\Delta v}{v}&\equiv&S\frac{\Delta h}{h}=\frac{1}{2}S\frac{\Delta\alpha}{\alpha}\label{eq:S},\\
\frac{\Delta \Lambda_{QCD}}{\Lambda_{QCD}}&\equiv&R\frac{\Delta\alpha}{\alpha}\text{ + (EW terms)},\label{eq:R}
\eq
where $v$ represents the Higgs vacuum expectation value, $h$ the Yukawa couplings, $\Lambda_{QCD}$ the Quantum Chromodynamics (QCD) mass scale. The latter equality in Eq. (\ref{eq:S}) provides the relation between the relative variation in the Yukawa couplings and that of $\alpha$. Each pair of values $(R,S)$ phenomenologically represents one particular GUT model. Relative variations of various parameters can then be written in terms of these three parameters \cite{Martins1}. Starting with the electron, proton, and neutron masses,
\bq
\frac{\Delta m_e}{m_e}&=&\frac{1}{2}\left(1+S\right)\frac{\Delta\alpha}{\alpha},\label{eq:dme}\\
\frac{\Delta m_p}{m_p}&=&\left[0.2\left(1+S\right)+0.8R\right]\frac{\Delta\alpha}{\alpha}\,,\label{eq:dmp}
\eq
while for the mass difference between protons and neutrons, $Q_N$,
\be
\frac{\Delta Q_N}{Q_N}=\left[0.1+0.7S-0.6R\right]\frac{\Delta\alpha}{\alpha},\label{eq:dQN}
\ee
and therefore
\be
\frac{\Delta m_n}{m_n}=\frac{\Delta Q_N}{Q_N}+\frac{m_p}{m_n}\left(\frac{\Delta m_p}{m_p}-\frac{\Delta Q_N}{Q_N}\right)\,.\label{eq:dmn-dmp-dQN}
\ee
Another impacted quantity is the neutron lifetime $\tau_n$,
\be
\frac{\Delta\tau_n}{\tau_n}=\left[-0.2-2S+3.8R\right]\frac{\Delta\alpha}{\alpha}\,.\label{eq:dtaun}
\ee

The alternative approach, which we treat separately in the following sections, is varying $G_N$ without any variations of particle masses. Arguably, this choice is less well motivated in a GUT models context, but considering that all this approach is necessarily phenomenological, this alternative approach is useful precisely as a comparison point. In this case the proton and electron masses do not vary (i.e. Eqs.(\ref{eq:dme}--\ref{eq:dmn-dmp-dQN}) do not apply) and instead, for the same relative variation of the observable $\alpha_g$, we have
\be
\frac{\Delta G_N}{G_N}=\left[0.4\left(1+S\right)+1.6R\right]\frac{\Delta\alpha}{\alpha}\,,\label{eq:dGN}
\ee
For purely gravitational processes the two choices should be observationally equivalent, but, as has been emphasized already, in BBN gravitational effects compete with non-gravitational ones, and therefore the approaches two will lead to somewhat different results, e.g. varying masses impact other sectors of the model. As a caveat, we note that Eq.(\ref{eq:dGN}) specifically depends on the fact that the proton mass is used in the definition of $\alpha_g$, and its coefficients would change if instead one defined $\alpha_g$ using the neutron or electron masses.

For $G_F$ a general expression is \cite{Coc}
\be
\frac{\Delta G_F}{G_F}=-S \frac{\Delta\alpha}{\alpha}\,.\label{eq:dGF-dalpha}
\ee
In the varying $G_N$ case we could also use the relation \cite{Dent}
\be
\frac{\Delta\tau_n}{\tau_n}=-2\frac{\Delta G_F}{G_F}\,,\label{eq:dGF-dtaun}
\ee
which can straightforwardly be introduced in the code. Since an equality works in both directions, and in any case these relative variations are always included in the code as functions of $\Delta\alpha/\alpha$, $R$ and $S$, one could in principle invert Eq. (\ref{eq:dGF-dtaun}) and use it as a numerical alternative to the general expression given by Eq. (\ref{eq:dGF-dalpha}) for the relative variation of $G_F$. One can numerically check that the choice of which of the two expressions is used in the code has a much smaller impact on the results that the previously introduced choice of varying $G_N$ versus varying masses.

One also needs the variations for the anomalous magnetic moments of the proton and the neutron, $\kappa_p$ and $\kappa_n$ respectively. They can be found from the expressions for the variation of the gyromagnetic ratios $g_p$ and $g_n$ \cite{kappas}
\bq
\frac{\Delta g_p}{g_p}&=&\left[0.1R-0.04\left(1+S\right)\right]\frac{\Delta\alpha}{\alpha},\label{eq:dgp for kappas}\\
\frac{\Delta g_n}{g_n}&=&\left[0.12R-0.05\left(1+S\right)\right]\frac{\Delta\alpha}{\alpha},\label{eq:dgn for kappas}
\eq
noting that $\kappa_p=g_p/2-1$ and $\kappa_n=g_n/2$. Finally, the W and Z particle masses can be written
\bq
m_W &=& \frac{1}{2}g_2 v,\label{eq:mW}\\
m_Z &=& \frac{1}{2}\sqrt{g_1^2+g_2^2} v,\label{eq:mZ}
\eq
where $g_2$ and $g_1$ are the SU(2) and U(1) gauge couplings and $v$ is the Higgs vacuum expectation value as in Eq.(\ref{eq:S}). Assuming, as in \cite{Coc}, that the relative variations of all the gauge and Yukawa couplings are the same, $\Delta g_i/g_i=\Delta g/g=1/2 \Delta\alpha/\alpha$, and that $\Delta v/v=S\Delta g/g$ from Eq.(\ref{eq:S}), we have
\bq
\frac{\Delta m_W}{m_W}&=&\frac{1}{2}(1+S)\frac{\Delta\alpha}{\alpha},\label{eq:dmW}\\
\frac{\Delta m_Z}{m_Z}&=&\frac{1}{2}(2+S)\frac{\Delta\alpha}{\alpha}\,.\label{eq:dmZ}
\eq

Our baseline model is the standard one, with $\Delta\alpha/\alpha=0$, in which case the values of $R$ and $S$ are irrelevant. In general ($\Delta\alpha/\alpha$, $R$, $S$) will be taken as free parameters, with ample top-hat priors. Nevertheless, it is useful, for illustration purposes, to have some specific choices of $(R,S)$ which may be representative of the allowed parameter space. For this purpose we consider four such choices, all of which have been previously considered in the literature: ($36$, $160$) \cite{Coc}, ($109.4$, $0$) \cite{Nakashima}, ($-183$, $22.5$) \cite{Rnegativo}, and ($0$, $-1$) \cite{Martins1}. The first is seen as a 'typical' GUT model; the second is a dilaton-type model which may be illustrative of string theory; the third has the peculiarity of having a negative $R$ (which is more commonly positive); finally, the fourth is a limiting case (or minimal model) where $\alpha$ still varies but several other quantities do not.

\section{\label{section3}Numerical implementation and tests}

The possible variations of fundamental couplings allowed by this class of GUT models and described in the previous section were implemented in the {\tt PRyMordial} code \cite{Burns}. A brief summary, focusing on the list of the impacted code variables, can be found in Appendix \ref{app1}; further details on the numerical implementation, as well as additional tests, can be found in \cite{Thesis}. In addition to the changes in the impacted variables, specific flags allow a choice between the two options for the gravitational sector (varying $G_N$ or varying masses) and for the variation of $G_F$.

To validate our code, in addition to reproducing standard results in the case of no variations, we have compared it to a similar analysis done in \cite{Coc}, aiming to reproduce some of the plots therein. The results were satisfactory though not entirely identical, the obvious reason being that the nuclear reaction rates and cross sections used in the earlier work (which was published in 2007) were somewhat different from the more recent ones we use.

\begin{figure}
\centering
\includegraphics[width=\columnwidth]{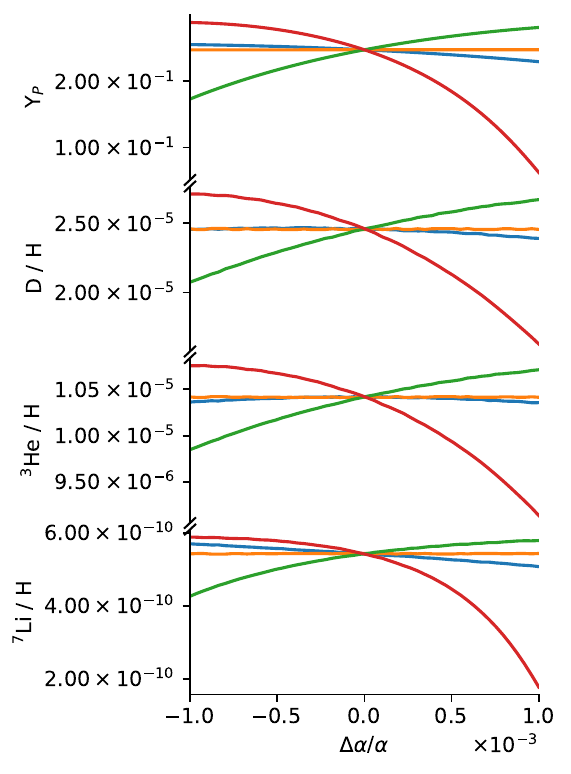}
\caption{Primordial abundances as a function of $\Delta\alpha/\alpha$ for the four representative (R,S) pairs discussed in the text: ($36$, $160$) in blue, ($109.4$, $0$) in green, ($-183$, $22.5$) in red, and ($0$, $-1$) in orange. Varying masses have been assumed in the gravitational sector, and $Y_p$ is the Helium-4 mass fraction.}
\label{fig01}
\end{figure}

Figure \ref{fig01} shows an example of the code outputs: the abundances of the four main nuclei are plotted as a function of $\Delta\alpha/\alpha$ for the four representative models mentioned at the end of the previous section. In the gravitational sector masses were allowed to vary, and for $G_F$ the general expression, Eq. (\ref{eq:dGF-dalpha}), has been assumed.

We have observed that simulation failures can sometimes occur for numerical reasons, as certain intermediate values in the calculation become so large that they are considered numerically infinite by Python. In practical terms this is not an issue, since the corresponding parameter values can be expected to not be viable models if the simulation has to work with unreasonable numbers. We also note that the numerical calculation includes some systematic errors, due to numerical round-offs, some of which happen only once (e.g. some data tables have single precision, that count as rounded values for double precision Python calculations), while others involve multiple round-offs as every step is a whole new simulation where an underlying variable is changed and every calculation is repeated with small round-offs in each.

\begin{figure}
\centering
\includegraphics[width=\columnwidth]{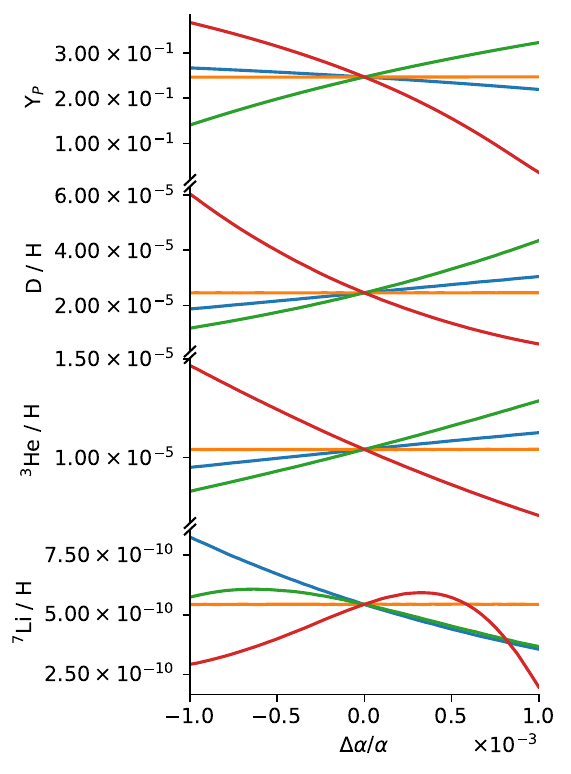}
\caption{Same as Fig. \ref{fig01}, but assuming a varying $G_N$ in the gravitational sector.}
\label{fig02}
\end{figure}
\begin{figure}
\centering
\includegraphics[width=\columnwidth]{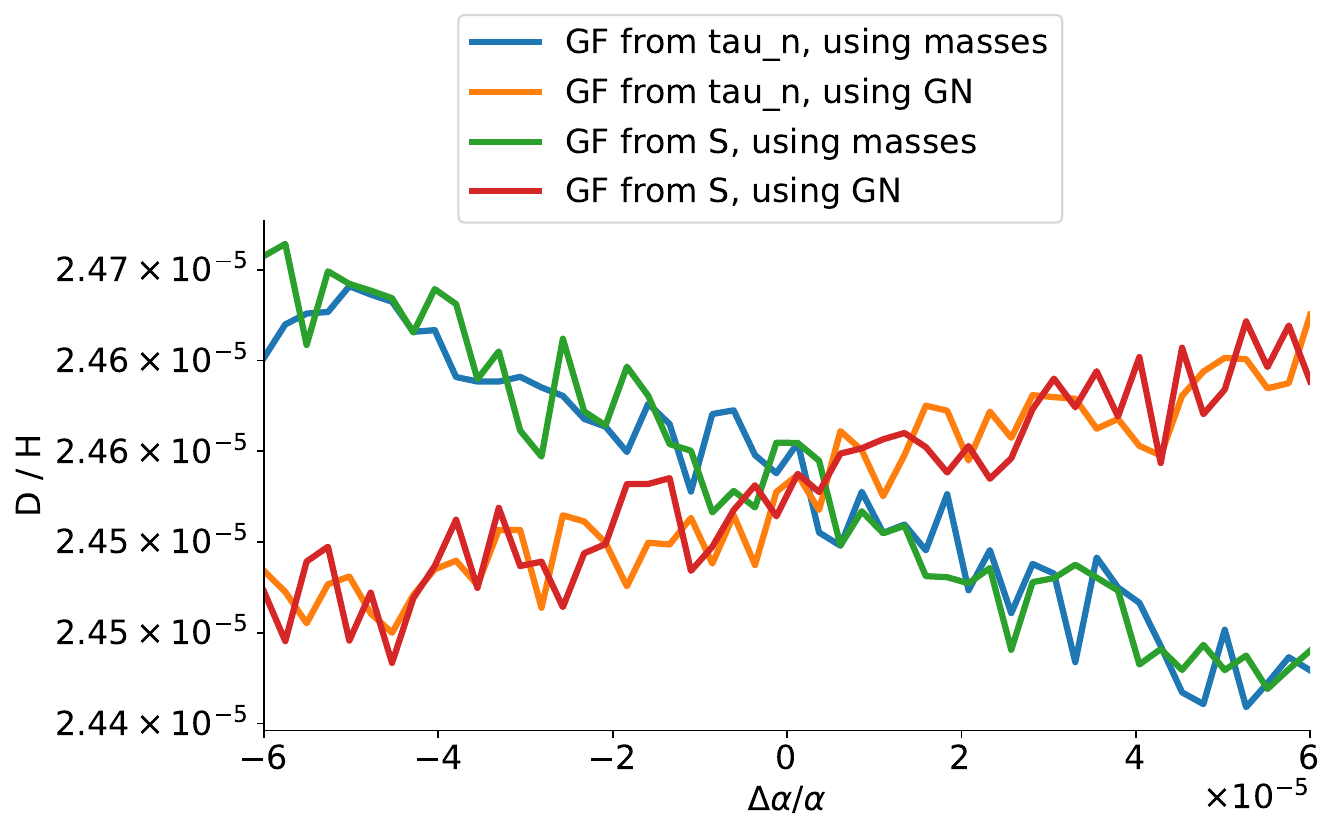}
\includegraphics[width=\columnwidth]{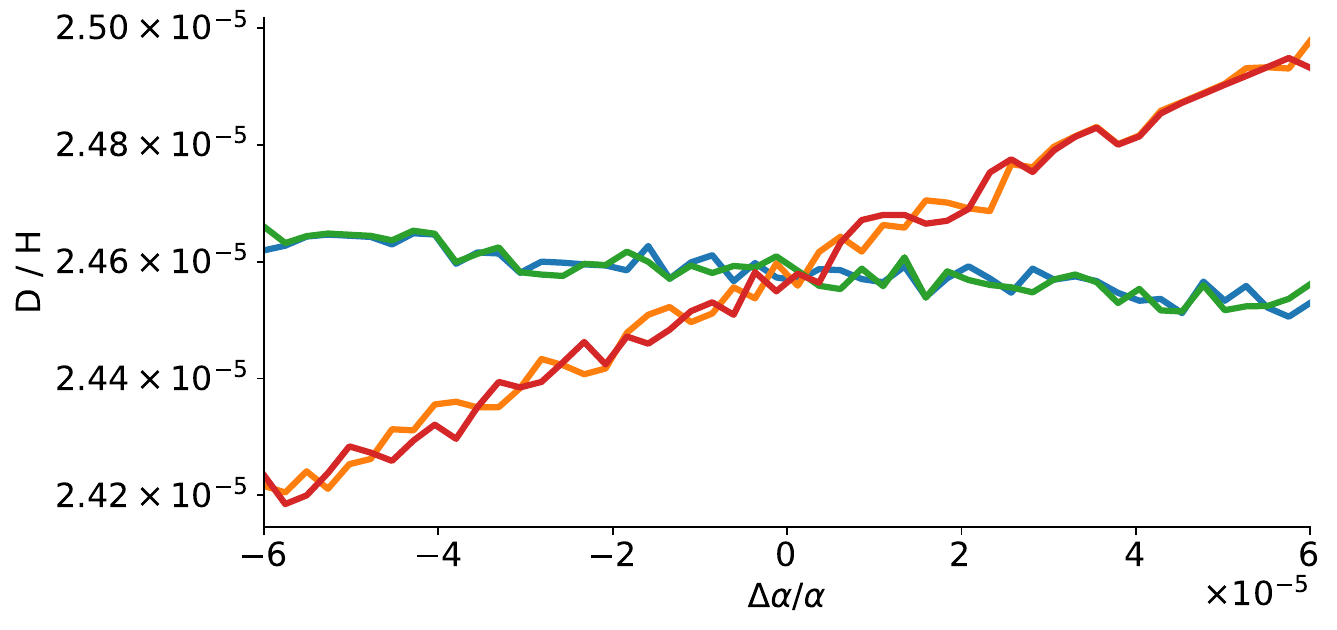}
\includegraphics[width=\columnwidth]{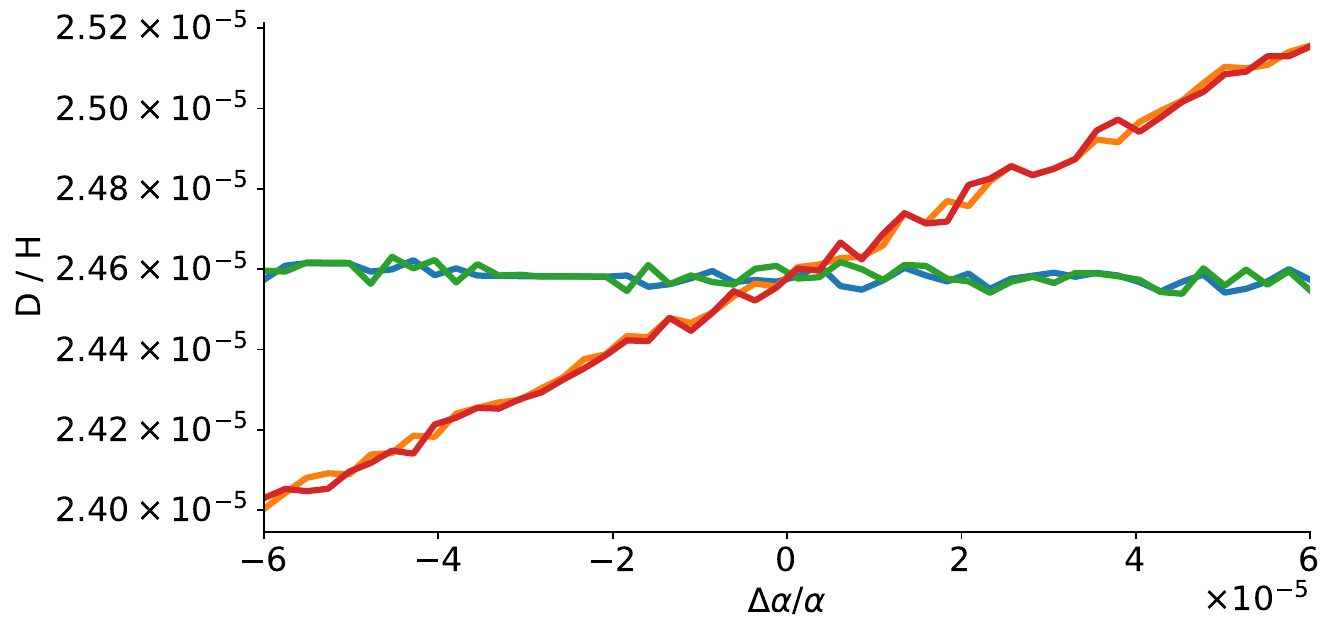}
\caption{Primordial Deuterium abundances as a function of $\Delta\alpha/\alpha$ for the four possible modeling choices for the gravitational sector and $G_F$. The value of $R$ is 0, 36 and 60 in the top, middle and bottom panels respectively, while $S=240$ in all cases.}
\label{fig03}
\end{figure}

The effect of varying Newton's constant instead of varying the masses is quite significant, as can be seen in Fig. \ref{fig02}, highlighting the fact that BBN is a competition between particle/nuclear physics and cosmology. This is to be expected: a variation of $G_N$ impacts the gravitational part of the code in a way that differs from the effect of mass variations. As for $G_F$, we have checked that the impact of using Eq. (\ref{eq:dGF-dtaun}) instead of Eq. (\ref{eq:dGF-dalpha})---both for the case of varying $G_N$, where the former equation holds, but also, as a numerical check, for the case of varying masses. In both cases, the impact of this choice is negligible: differences would not be discernible by eye in an analogous plot, This is further illustrated, for the Deuterium abundance, in Fig. \ref{fig03}, which also shows the impact of varying $R$ (while $S$ is kept constant). In these circumstances, we will keep studying the two alternative choices for the gravitational sector (varying masses and varying $G_N$).

Figure \ref{fig03}---which uses a higher resolution grid than those used in the previous figures--- also illustrates the aforementioned numerical uncertainties, in particular showing that they depend on the model parameters considered. Despite this, the time needed to run the code for one specific model is almost independent of the choice of model parameters, being about 20 seconds on a standard recent laptop. Note that for our purposes we need to recompute the background for each of our models (while in more standard scenario this background can often be pre-computed once and subsequently stored).

\section{\label{section4}Comparing with observations}

Our data analysis will rely on the estimated primordial abundance of Deuterium and Helium-4, for which there are robust cosmological measurements \cite{PDG,Cooke}. For Helium-3 the inferred abundances are local rather than cosmological (so they are not relevant for our analysis), while Lithium-7 is plagued by the well-known Lithium problem \cite{Lithium,PDG}, to which we will return later in this work. We use the observational abundances recommended in the latest available (2024) PDG review \cite{PDG},
\bq
Y_p&=&0.245\pm0.003\,,\\
\frac{D}{H}&=&(25.47\pm0.29)\times10^{-6}\,,\\
\frac{{}^7Li}{H}&=&(1.6\pm0.3)\times10^{-10}\,;
\eq
strictly speaking, the Lithium-7 abundance may be considered a lower bound.

For an initial comparison of the abundances calculated for different model parameters  with the observed data we draw grid-based charts, for fixed values of $\Delta\alpha/\alpha$, which span a range of values of $R$ and $S$. For each $(R,S)$ pair, a color quantifies the agreement (or lack thereof) between the model simulation and data best fits: green is within $3\sigma$, yellow is between $3\sigma$ and $5\sigma$, and red is at more than $5\sigma$. The choice of these thresholds is somewhat arbitrary, but they are simply meant as an indication of which parameters might provide reasonable fits to the data, in preparation for a more thorough analysis in the next section. To be able to compare different scenarios, we can draw each point as a circle divided into two or even four slices, each of them representing a different scenario; we therefore colloquially refer to these plots as 'cheese charts'.

As expected the value of $\alpha$ is crucial for the derived primordial abundance, so it is important to identify reasonable ranges for it. As will be clear in what follows, values of $\Delta\alpha/\alpha$ of $10^{-4}$ or smaller (in absolute value), are possibly allowed, while larger values are excluded, except if there is some fine-tuning of the $R$ and $S$ values. We also note that an approximate symmetry can be expected in the analysis, and seen in these plots. Since, for generic quantities $Q$, the sensitivity coefficients have the general form $\Delta Q/Q=(aR+bS+c)\Delta\alpha/\alpha$, where $(a,b,c)$ are order unity (or possibly zero) constants, simultaneously changing the signs of all three parameters ($\Delta\alpha/\alpha,R,S$) leads to a model which in terms of observable consequence is almost identical.

\begin{figure}
\centering
\includegraphics[width=\columnwidth]{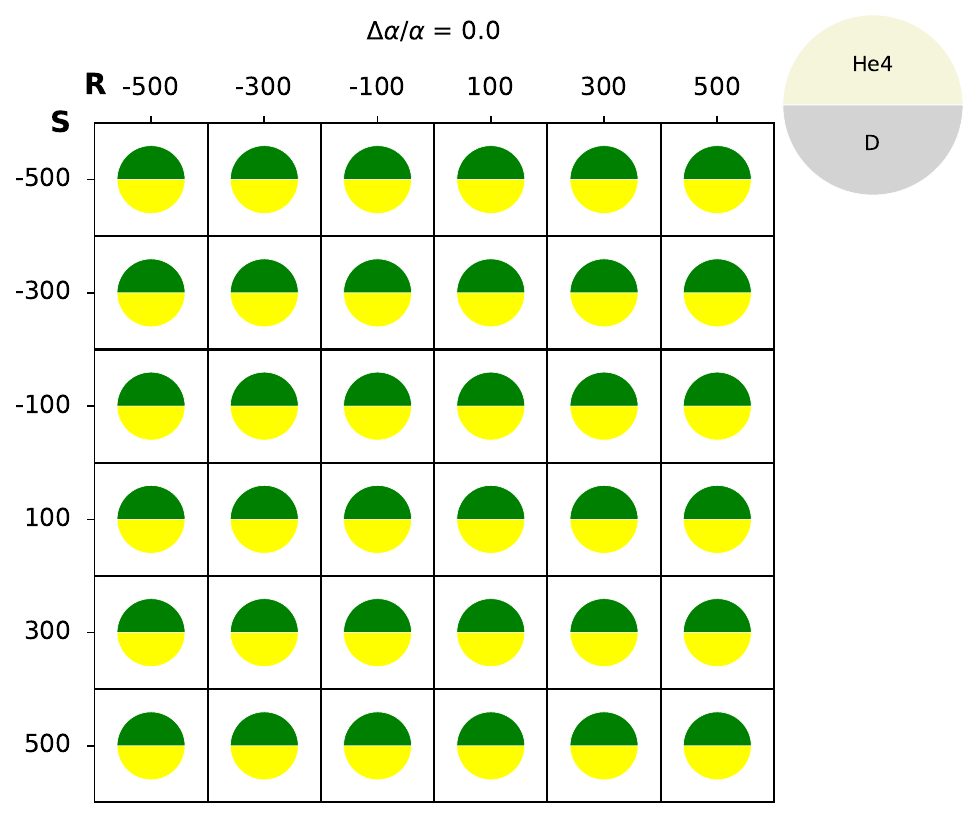}
\includegraphics[width=\columnwidth]{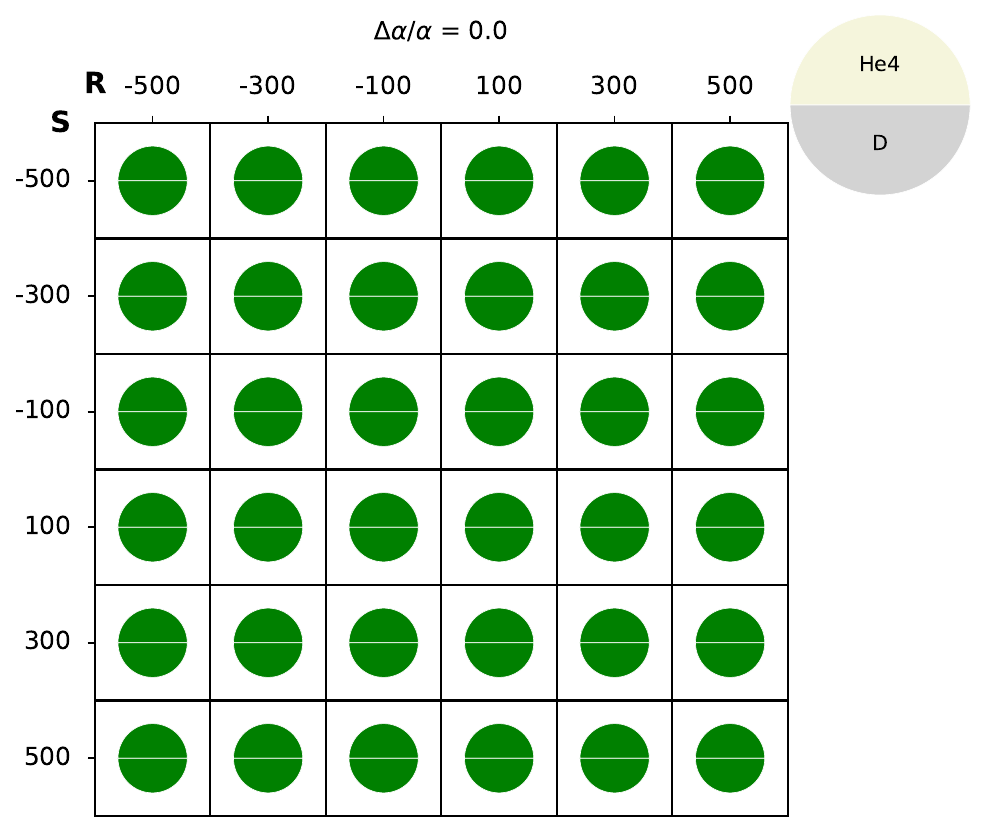}
\caption{Cheese chart for $\Delta\alpha/\alpha=0$ showing Helium-4 on top and
Deuterium on the bottom. Te top and bottom panels use the nuclear rates from the {\tt PRIMAT} and {\tt NACRE II} databases respectively.}
\label{fig04}
\end{figure}
\begin{figure}
\centering
\includegraphics[width=\columnwidth]{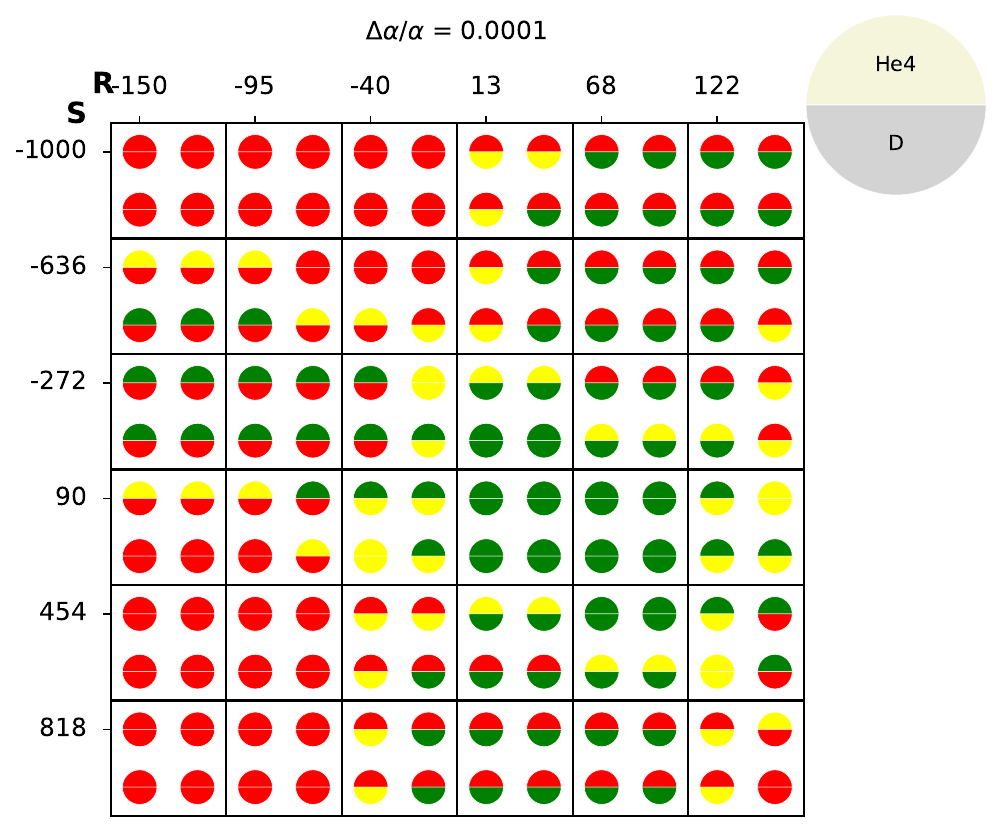}
\includegraphics[width=\columnwidth]{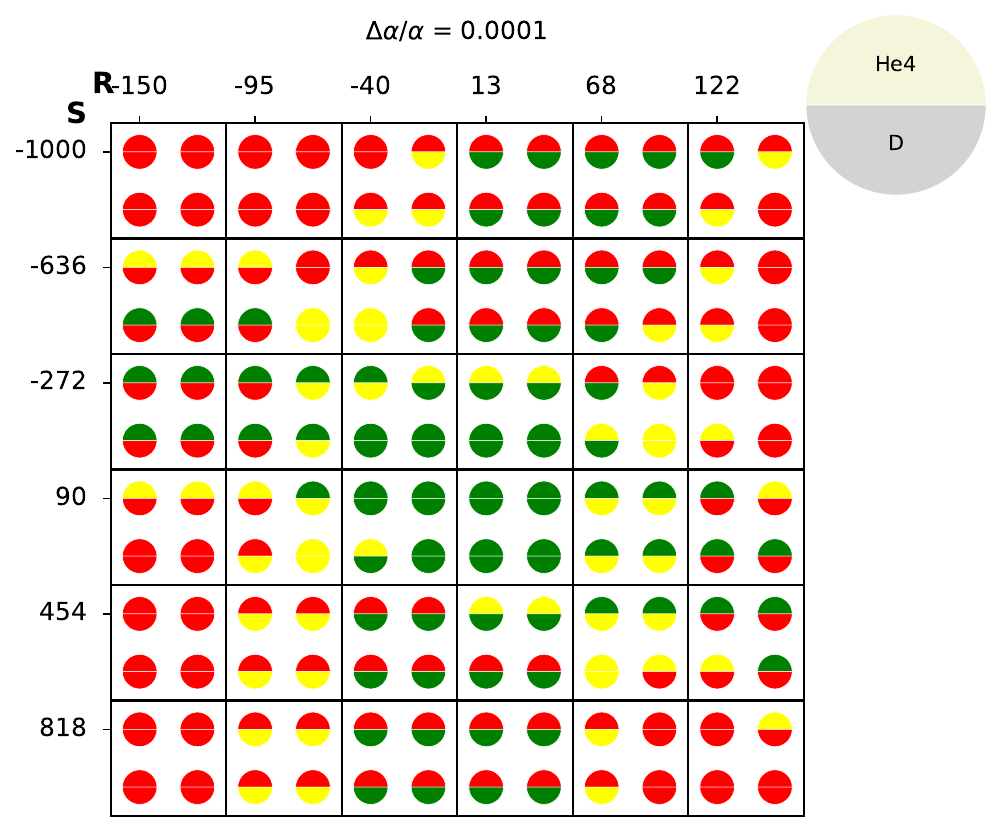}
\caption{Same as Fig. \ref{fig04}, for $\Delta\alpha/\alpha=+10^{-4}$. In the gravitational sector $G_N$ is assumed to vary,}
\label{fig05}
\end{figure}

Figure \ref{fig04} shows one example, for the case of no $\alpha$ variations, comparing the impact of relying on the nuclear rates from the databases of {\tt PRIMAT} \cite{primatHe4} and {\tt NACRE II} \cite{nacre} respectively. The top panel highlights the Deuterium discrepancy, which has been previously reported for {\tt PRIMAT} \cite{primatD} and also considered in the context of GUT models in \cite{Martins3}. Figure \ref{fig05} shows the same comparison for a grid of $(R,S)$ values with  $\Delta\alpha/\alpha=+10^{-4}$ and a gravitational sector with varying $G_N$.

\begin{figure}
\centering
\includegraphics[width=\columnwidth]{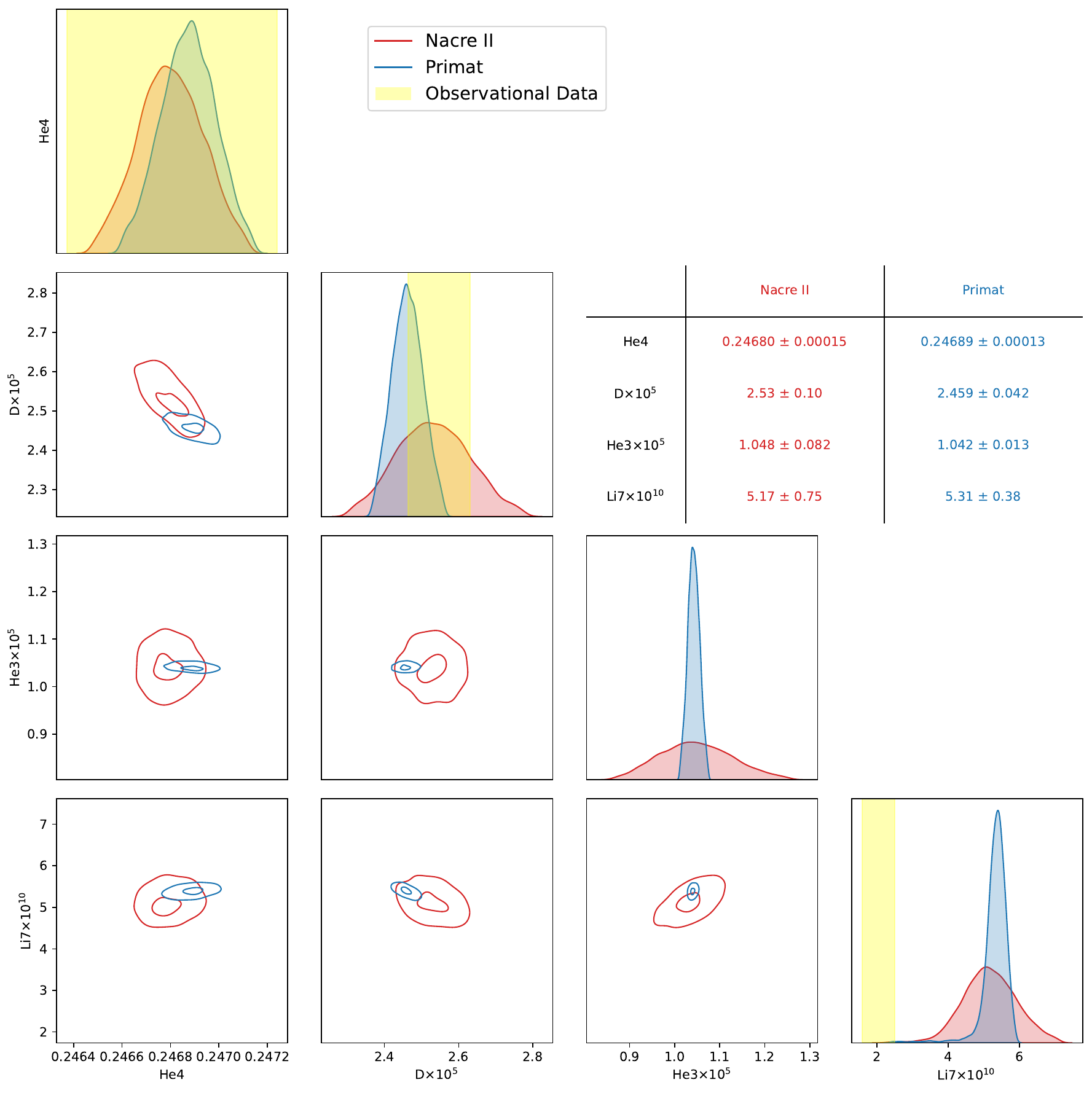}
\caption{The 1D probability distributions and 2D joint 68\% and 95\% probability regions for the four BBN abundances using {\tt NACRE II} (red lines) and {\tt PRIMAT} (blue lines), compared to the $3\sigma$ observational data ranges. This reproduces Figure 3 of \cite{Burns}.}
\label{fig06}
\end{figure}

Since our goal is to constrain deviations from the standard cosmological paradigm, in what follows we will conservatively use the {\tt NACRE II} database as our baseline, to minimize possible nuclear physics impacts. Although the {\tt PRIMAT} database is clearly more precise than {\tt NACRE II}, it is possible that it is less accurate. This can also be seen in Fig. \ref{fig06}, which reproduces Fig. 3 of \cite{Burns}, and is also a validation test of our code, for $\Delta\alpha/\alpha=0$. The plot uses the abundances in $10000$ steps with Gaussian priors for the baryon-to-photon ratio and the neutron lifetime and log-normal priors for the nuclear rates.

\begin{figure}
\centering
\includegraphics[width=\columnwidth]{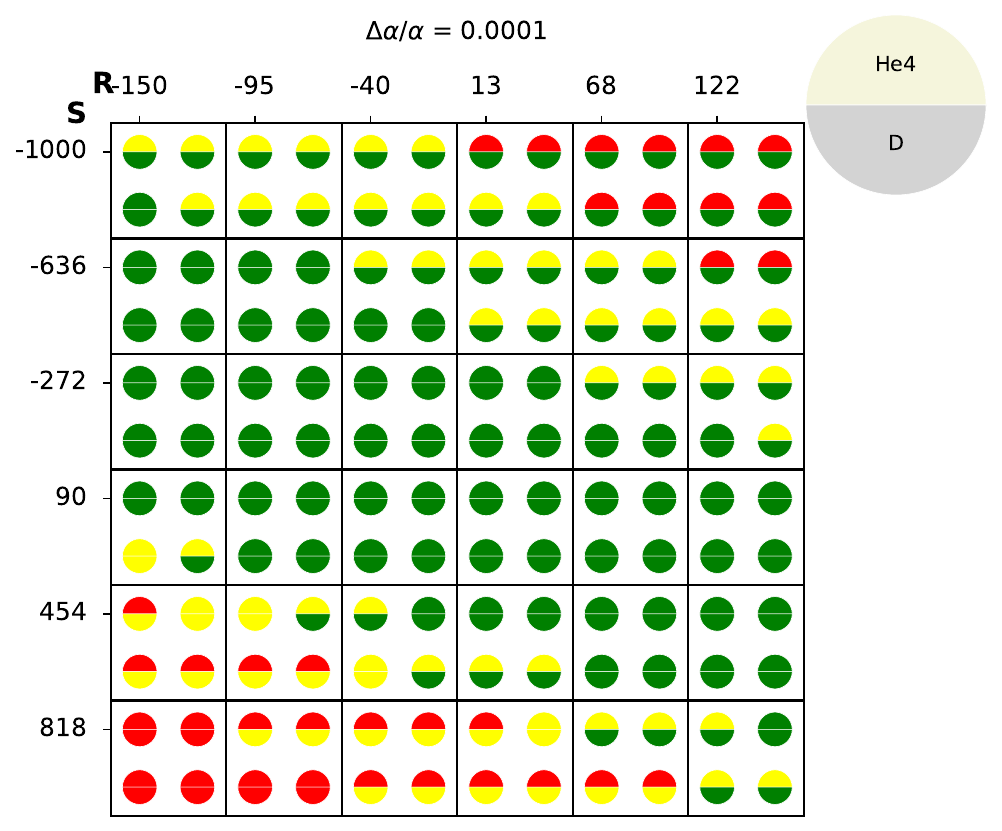}
\caption{Cheese chart for $\Delta\alpha/\alpha=+10^{-4}$ using the nuclear rates from the {\tt NACRE II} database. In the gravitational sector particle masses are assumed to vary,}
\label{fig07}
\end{figure}

By contrast, Fig. \ref{fig07} shows an analogous cheese chart using the nuclear rates from the {\tt NACRE II} database but allowing particle masses (rather than $G_N$) to vary. This is to be compared to the bottom panel of Fig. \ref{fig05}. It is striking that when $G_N$ is allowed to vary, the Helium-4 and Deuterium are sensitive to different $(R,S)$ linear combinations, leading to green bands in this parameter space which are approximately orthogonal. Thus a joint analysis of the two abundances breaks this degeneracy. On the other hand, if the gravitational sector has varying masses, the sensitivities are almost the same, and the degeneracy is unbroken. For the value of $\Delta\alpha/\alpha=+10^{-4}$ shown in Fig. \ref{fig07} the degeneracy direction in this plane is found to be $S\sim3.3R\pm454$, with the error bar representing the width of the green band. However, this is only an approximate value, and moreover it (especially the width) is expected to depend on the value of $\Delta\alpha/\alpha$. A more accurate value will be provided below.

\begin{figure}
\centering
\includegraphics[width=\columnwidth]{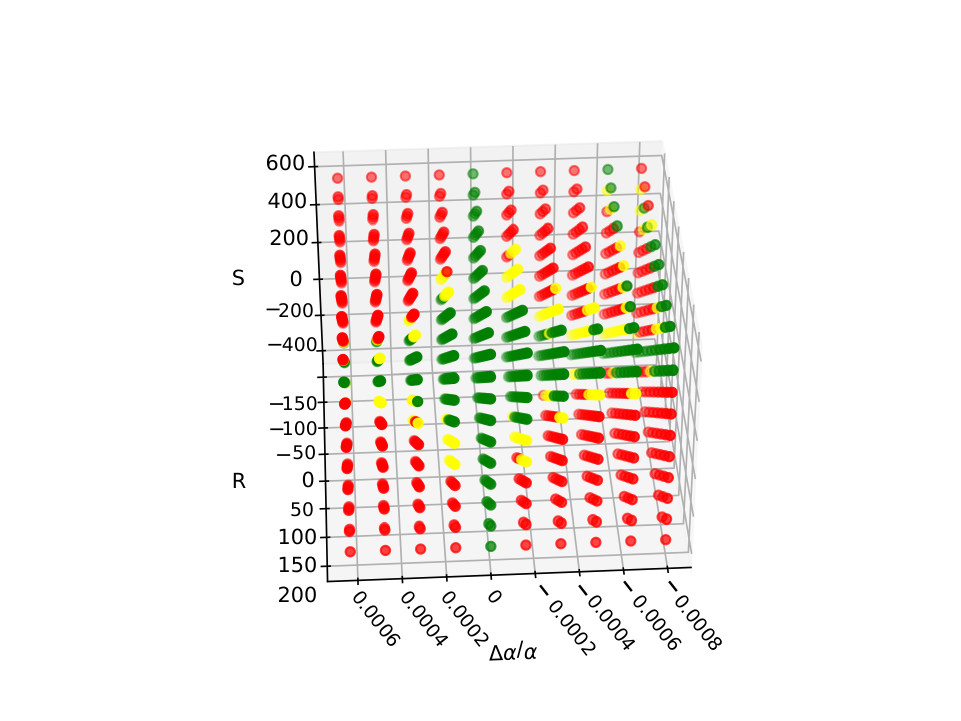}
\includegraphics[width=\columnwidth]{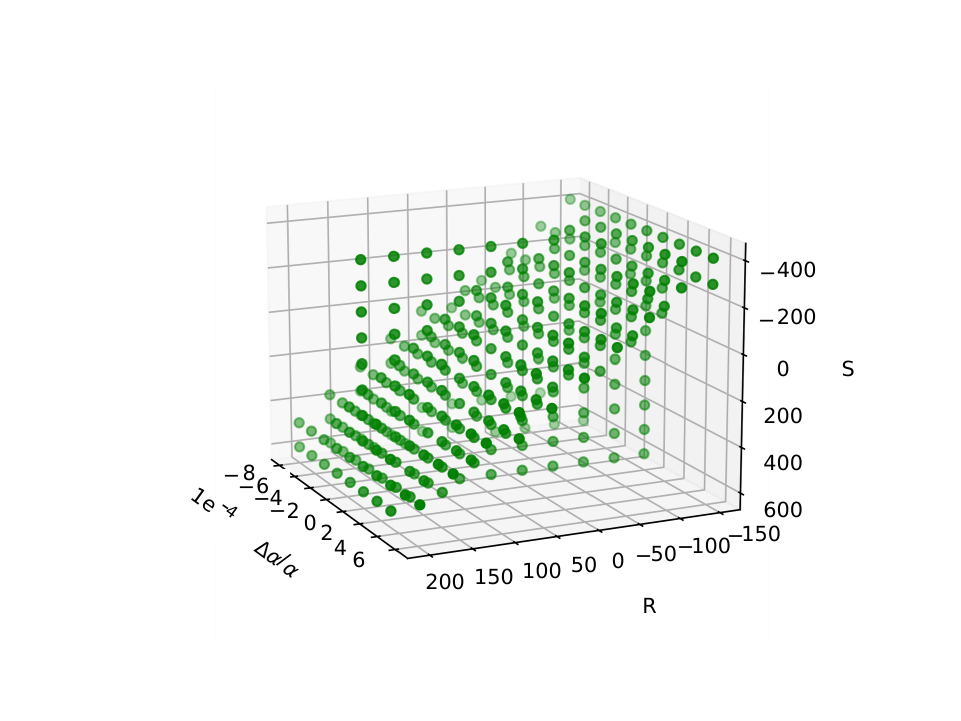}
\caption{Static scatter plot for the Helium-4 abundance, with varying masses in the gravitational sector. The top panel shows all thee color-coded regions, while the bottom panel only shows (from a different viewing point) the green region.}
\label{fig08}
\end{figure}

The two-dimensional cheese chart analysis can in principle be extended to three dimensions by also scanning along the $\Delta\alpha/\alpha$ direction, although visualizing such plots is not trivial. Figure \ref{fig08} shows one example \footnote{The interactive 3D
versions of these and other images can be found in \href{https://zenodo.org/records/17295823}{https://zenodo.org/records/17295823}.}, for the Helium-4 abundance, with varying masses in the gravitational sector. This figure makes it clear that the observationally preferred green region is the intersection of two planes: one at $\Delta\alpha/\alpha=0$ and one spanning a diagonal between $R$ and $S$. Taking a weighted average (weighted by the regression coefficient) of a reasonable range of $\alpha$ variations we obtain a slope
\be\label{slope}
\frac{S}{R}=2.53\pm0.06\,,
\ee
where the uncertainty is conservatively taken from the largest distance to this average, which we expect to be a more robust estimate than the one given earlier.

\section{\label{section5}Constraints on GUT models}

We can now obtain observational constraints on the ($\Delta\alpha/\alpha,R,S$) parameter space. As was mentioned in the context of Fig. \ref{fig06}, the standard {\tt PRyMordial} code assumes Gaussian priors for the baryon-to-photon ratio and the neutron lifetime and log-normal priors for the nuclear rates. In our case we have three additional free parameters, implying that a brute-force Monte Carlo exploration of the full parameter space would be computationally very long (except perhaps in a parallel version of the code). However, there are more efficient ways to carry out the analysis.

To begin with, one should understand where the main source of uncertainties lies. By running the standard code while giving priors for the baryon-to-photon ratio, neutron lifetime the nuclear rates one at a time, it becomes clear that the uncertainty related to $\tau_n$ is comparatively very small and will carry little significance. Clearly the most significant source of uncertainty comes from the nuclear rates, which is unsurprising given the differences {\tt PRIMAT} and {\tt NACRE II}.

\begin{table}
    \centering
	\begin{tabular}{|c|ccccccc|} 
		\hline
		$\Delta\alpha/\alpha$ & $-10^{-3}$ & $-10^{-4}$ & $-10^{-5}$ & 0 & $10^{-5}$ & $10^{-4}$ & $10^{-3}$\\
		\hline
		 ${}^4He\times10^4$ & 1.33 & 1.53 & 1.49 & 1.49 & 1.52 & 1.56 & 1.81 \\
      $D\times10^6$ & 1.01 & 1.03 & 1.02 & 0.97 & 1.01 & 1.04 & 0.97 \\
		\hline
	\end{tabular}
	\caption{Standard deviations of distributions for different values of $\Delta\alpha/\alpha$, due to the uncertainties in the nuclear reaction rates. Note that they have been multiplied by large factors (they are intrinsically quite small).}
	\label{tab01}
\end{table}

Next, there is the question of whether (and, if so, how) this uncertainty changes for different values of ($\Delta\alpha/\alpha,R,S$). With exactly the same assumptions on the distributions as before, we show in Table \ref{tab01}, as an example, the results for the standard deviations of these distributions for various choices of $\Delta\alpha/\alpha$, for constant $R=36$ and $S=60$. These results, as well as an analogous ones for different values of $R$ and $S$, show that the impact of these parameters, within the range of $\alpha$ variations that is observationally plausible, is very small. In other words, assuming that these standard deviations are constant (i.e., independent of the specific GUT model parameters) is a good approximation. Nonetheless, one can further sample this parameter space and derive linear and even quadratic fits for these small dependencies. This analysis is reported in \cite{Thesis}, which also shows that, to at least two decimal places, the standard deviations of these distributions are the same whether one is assuming varying masses or varying $G_N$.

\begin{figure}
\centering
\includegraphics[width=\columnwidth]{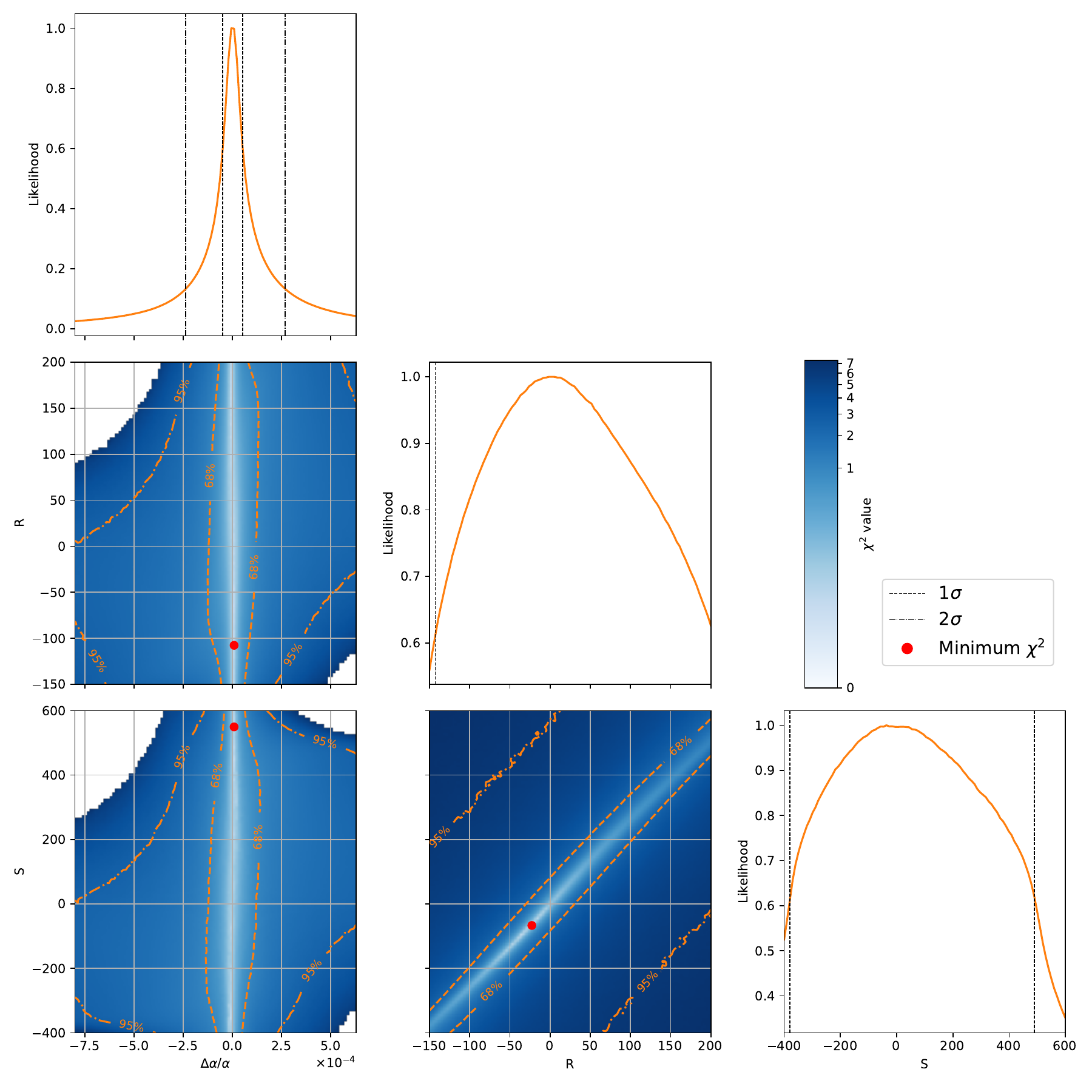}
\includegraphics[width=\columnwidth]{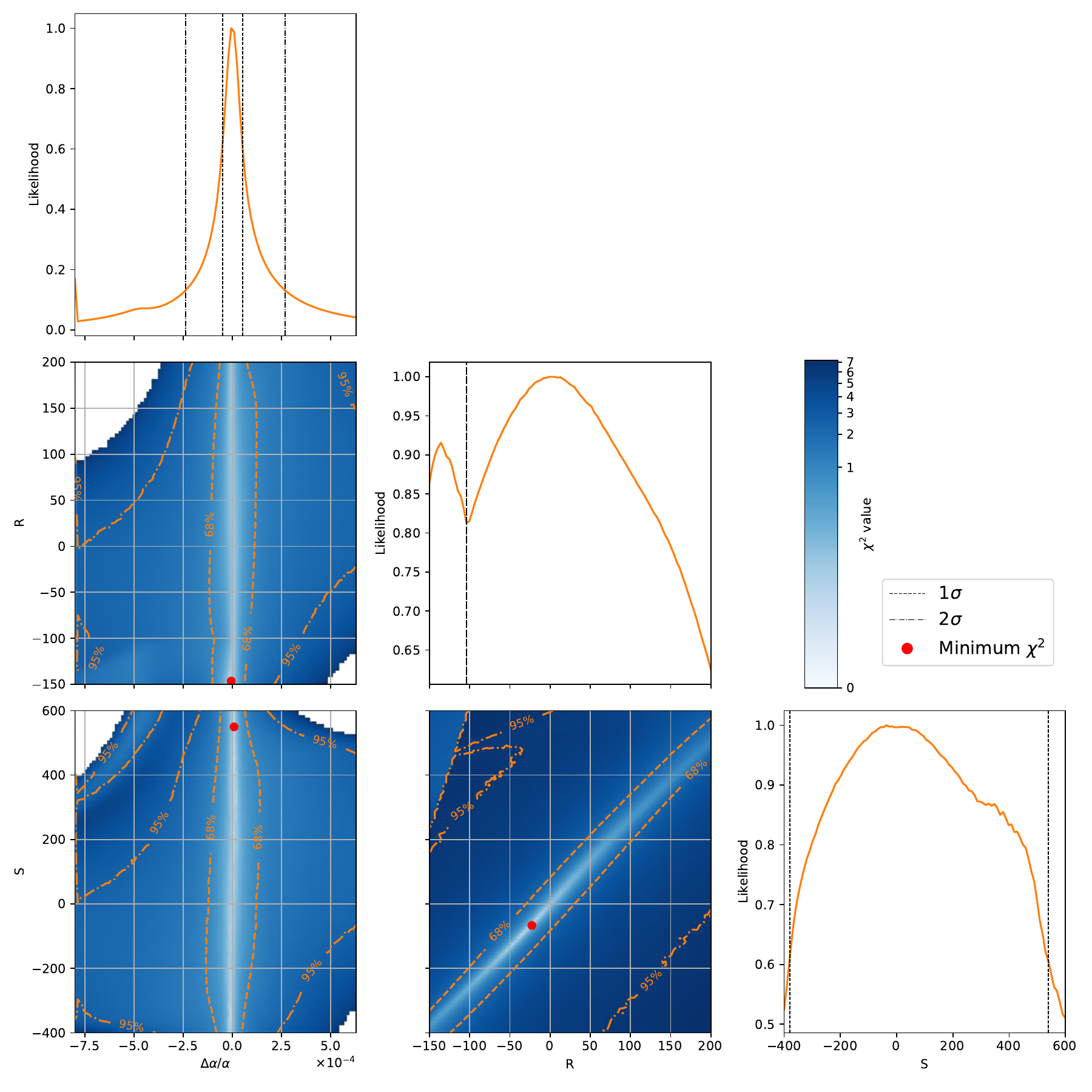}
\caption{BBN constraints on GUT models, for varying masses. In the top panel the analysis includes only the observational uncertainties, while in the bottom one both the observational and simulation uncertainties are included. The colormap in 2D panels shows the total $\chi^2$, but values exceeding Python's numerical capabilities are not plotted.}
\label{fig09}
\end{figure}
\begin{figure}
\centering
\includegraphics[width=\columnwidth]{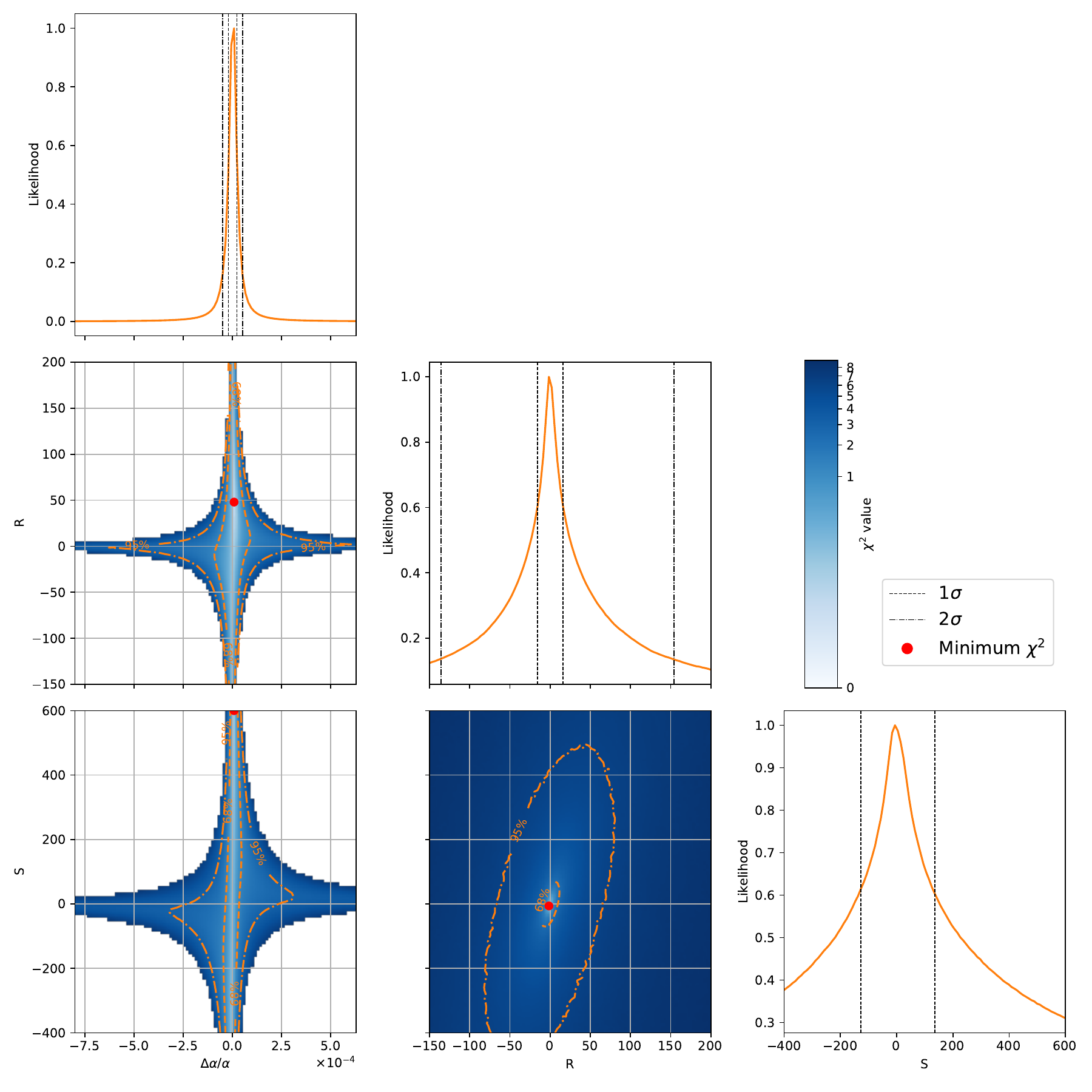}
\includegraphics[width=\columnwidth]{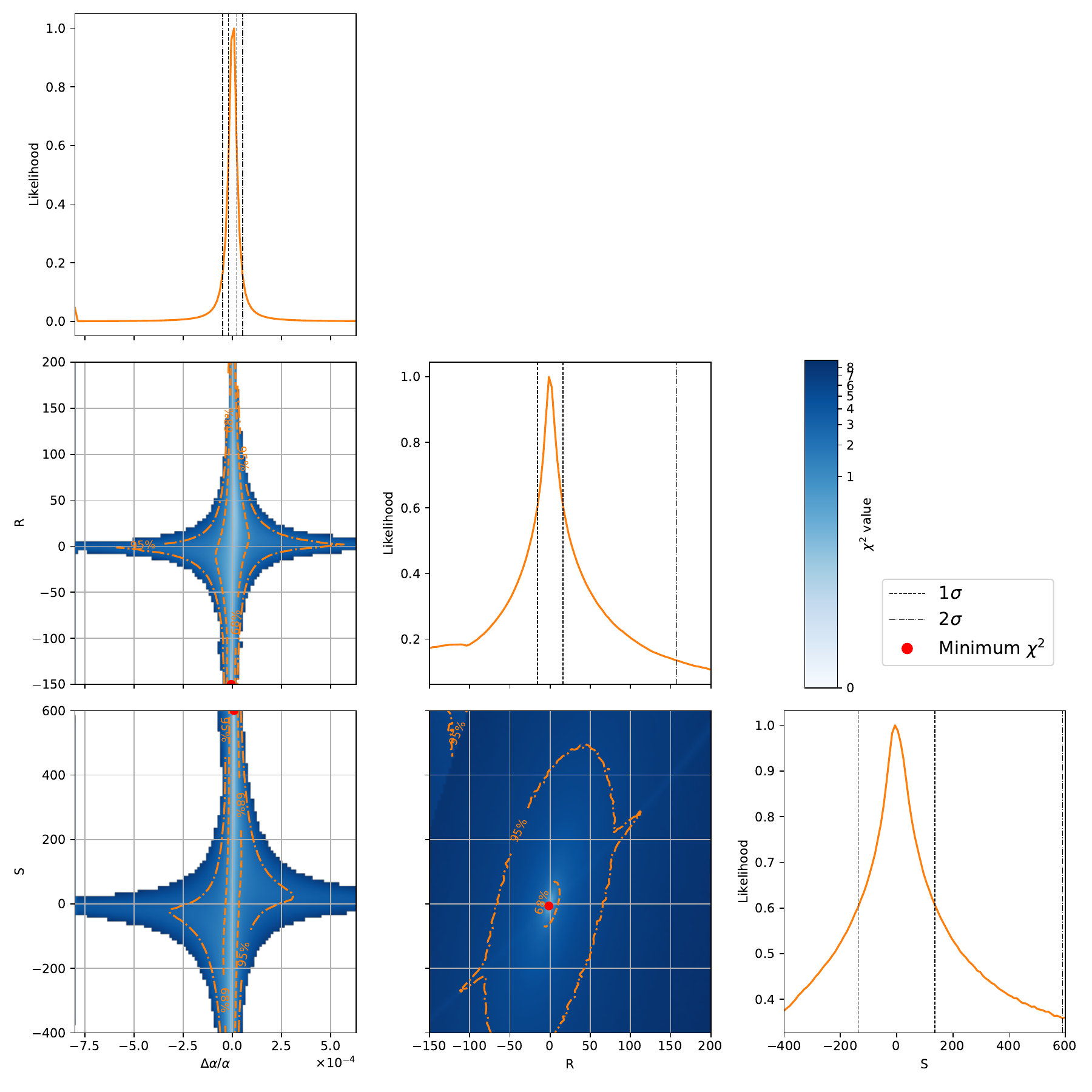}
\caption{Same as Fig. \ref{fig09}, for the case of varying $G_N$.}
\label{fig10}
\end{figure}

With these approximations for the uncertainties, we can carry out a maximum likelihood analysis in our 3-dimensional ($\Delta\alpha/\alpha,R,S$) space without needing to simulate $1000$ times in each point to get an uncertainty; instead, we use a pre-computed simulation uncertainty at each point. As previously mentioned, our analysis will rely on the estimated primordial abundance of Deuterium and Helium-4, so our chi-square has the form
\be
\chi^2=\frac{(He_{sim}-He_{obs})^2}{(\sigma^{He}_{sim})^2+(\sigma^{He}_{obs})^2}+\frac{(D_{sim}-D_{obs})^2}{(\sigma^{D}_{sim})^2+(\sigma^{D}_{obs})^2}\,.\label{eq:chi2}
\ee
The results of this analysis are shown in Figs. \ref{fig09}--\ref{fig10} for the cases of varying masses and varying $G_N$ respectively. In both cases, the top panels show the results obtained if one includes only the observational uncertainties, while the bottom panels include  both the observational and simulation uncertainties. It is noticeable that the inclusion of the simulation uncertainties has no significant effects around the peak of the likelihood (which is consistent with the null result, and therefore corresponds to rather small variations of $\alpha$) but does have some impact on the likelihood tails.

For the variation of masses, we obtain a $68\%$ confidence level constraint
\be
\frac{\Delta\alpha}{\alpha}=2\pm51\, ppm\,;
\ee
moreover, we confirm the previously discussed degeneracy between $R$ and $S$, in this case with an estimated slope $S=2.59R$, which is consistent, within uncertainties, with our earlier estimate reported in Eq. (\ref{slope}). On the other hand, for the varying Newton's constant, we can separately constrain the three parameters
\bq
\frac{\Delta\alpha}{\alpha}&=&2\pm22\, ppm\\
R&=&-2_{14}^{+16}\\
S&=&-6_{-121}^{+141}\,;
\eq
note that these likelihoods are highly non-Gaussian. These plots, especially in the $G_N$ case, also highlight the previously mentioned multiplicative symmetry.

\section{\label{section6}Coda: The Lithium problem revisited}

The Lithium problem is a key open issue in astrophysical cosmology: simulations and observations don't agree on the amount of Lithium-7 that was formed during BBN \cite{Lithium}. This could either be the fault of the theory behind the simulations, of the experimental data which is necessary for this analysis (e.g., the nuclear reaction rates), of the methods of observation, or (arguably more likely) of unknown astrophysical causes of destruction that aren't being taken into account. 

In principle, one might expect that varying fundamental constants are a promising route to explore in this regard. Broadly speaking, for a given amount of $\alpha$ variation, one expects stronger impacts for heavier nuclei, and therefore one may think that for some of these models the variations are small enough to keep Helium-4 and (especially) Deuterium within observationally acceptable values while changing the Lithium-7 abundance by the factor of three or so which would be needed for an outright solution to the problem. A superficial look at Figs. \ref{fig01}--\ref{fig02} would seemingly confirm this suspicion.

\begin{figure}
\centering
\includegraphics[width=\columnwidth]{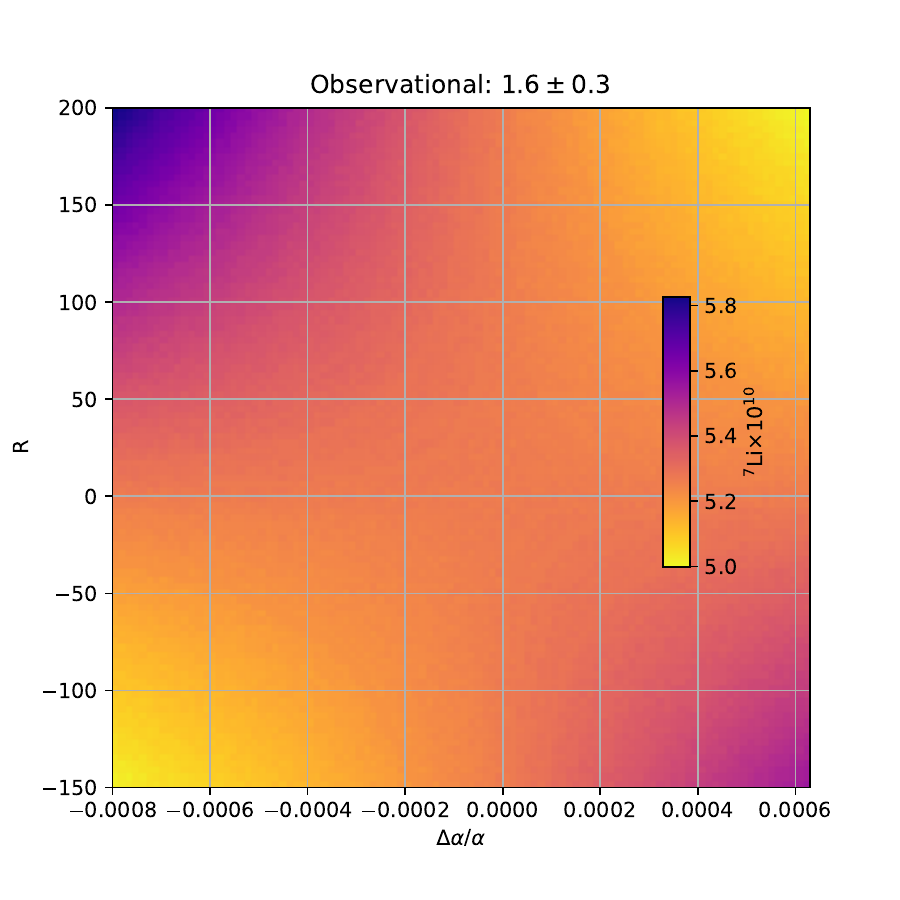}
\includegraphics[width=\columnwidth]{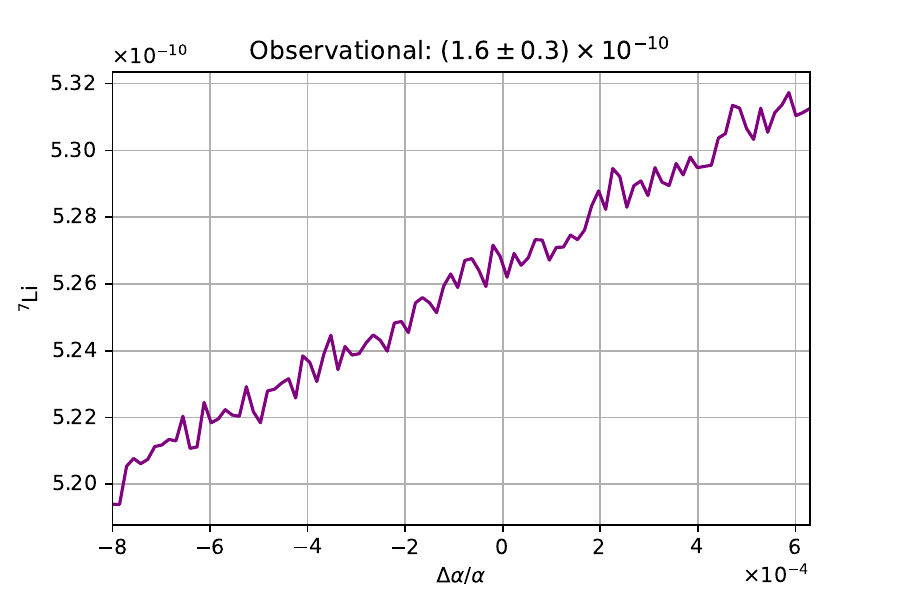}
\caption{Theoretically predicted Lithium-7 abundances, for models preferred by Deuterium and Helium-4 data. Top panel (varying masses): abundance in the 2-dimensional $(\Delta\alpha/\alpha,R)$ space, with $S$ set to $2.53R$. Bottom panel (varying $G_N$): abundance for $(R,S)=(-2,-6)$ and for different values of $\alpha$.}
\label{fig11}
\end{figure}

Unfortunately, as already suggested in \cite{Martins3}, a detailed analysis shows that no such goldilocks model exists: the variations of $\alpha$ allowed by the Deuterium and Helium-4 observations are such that the impact on the Lithium-7 abundance is necessarily small. Figure \ref{fig11} illustrates this point, for the GUT models preferred by the Deuterium and Helium-4 data, for both the varying masses and varying $G_N$ cases. Once can certainly find models for which the Lithium-7 abundance changes in the right direction---in other words, the theoretically predicted abundance is lowered---but these changes are too small to be a full solution. At most, these models could provide a subdominant contribution to a final answer which might be on the nuclear physics or astrophysics sides.

\section{\label{section7}Conclusions}

We have extended the {\tt PRyMordial} BBN code to provide self-consistent constraints on a broad class of GUT models in which there are multiple varying fundamental couplings. A previously developed self-consistent perturbative approach has been implemented in the code, and we have explored the impact of alternative assumptions on the weak and gravitational sectors, as well as that of the choice of nuclear reaction rates databases.

Overall, the constraints on $\alpha$ are at the tens of ppm level. They are slightly weaker than those reported in \cite{Martins1,Martins2}, which was based on analytic perturbative estimates. The main reason for this is that in our present numerical implementation further quantities are allowed to vary. The effects of these partially cancel each other, allowing for somewhat larger variations. The constraints are also weaker than those obtained at low-redshift from high-resolution spectroscopy along the line of sight of bright quasars, e.g. \cite{Murphy}, but they are of course at far higher redshifts---and much stronger than those obtainable from the cosmic microwave background.

One caveat of our analysis is that no explicit perturbations are included for nuclear reaction rates and binding energies, since we're not aware of any prescription that self-consistently applies to all those included in {\tt PRyMordial}. Considering that even in the standard case there are significant zeroth-order uncertainties in the nuclear reaction rates, as demonstrated by the differences between  {\tt PRIMAT} \cite{primatHe4} and {\tt NACRE II} \cite{nacre}, and that $\alpha$ variations are necessarily small (i.e. clearly in the perturbative regime) we believe that this simplification does not dominate our error budget. However, we note that this it is an assumption that warrants additional exploration.

Our analysis shows that BBN is a competitive probe of these unification scenarios. More precise measurements of primordial abundances, e.g. from the ELT \cite{ANDES}, should enable constraints at the ppm level. We also note that in this class of models $\alpha$ can be redshift-dependent but the parameters $(R,S)$ being phenomenological but constant parameters, applicable at all redshifts. It will therefore be interesting to combine this BBN analysis with those of other astrophysical data and local data which also enable constraints on these parameters, including atomic clocks, high-resolution astrophysical spectroscopy, and compact astrophysical objects such as white dwarfs and neutron stars \cite{other_RS_clocks,other_RS_compact}. These joint analyses are left for future work.
 
\appendix
\section{\label{app1}Further details on numerical implementation}

In what follows we list the variables in the {\tt PRyMordial} code impacted (or not) by our changes. Note that we do not list those variables that are written in the code as a function of others, since the variation of these will be automatically implemented when the underlying variables are varied. For example, {\tt QN=mn-mp} doesn't need to be changed directly, if {\tt mn} and {\tt mp} are already changed.

In {\tt PRyM\_init.py}, the following variables (using the variable names as given in the code) could be affected by our perturbative corrections
\begin{itemize}
    \item {\tt alphaem}: fine structure constant
    \item {\tt GF}: Fermi coupling constant
    \item {\tt mZ}: Z mass
    \item {\tt me}: electron mass
    \item {\tt mn}: neutron mass
    \item {\tt mp}: proton mass
    \item {\tt GN}: Newton constant
    \item {\tt gA}: axial current constant of structure
    \item {\tt kappa\_p}: anomalous magnetic moment of the proton
    \item {\tt kappa\_n}: anomalous magnetic moment of the neutron
    \item {\tt radproton}: proton charge radius
    \item {\tt tau\_n}: neutron lifetime
    \item {\tt Vud}: prediction for the CKM matrix entry from table I in \cite{Vud}  
\end{itemize}

In {\tt PRyM\_eval\_nTOp.py}, inside of the definition of the function {\tt RadCorrResum()}, with values taken from equations in \cite{RadCorrResum}, we have
\begin{itemize}
    \item {\tt mA}: axial vector meson mass from Eq.(9) in \cite{RadCorrResum}
    \item {\tt Agndecay}: from Eq.(9)in \cite{RadCorrResum}
    \item {\tt Cndecay}: from Eq.(9) in \cite{RadCorrResum}
    \item {\tt deltandecay}: from Eq.(12) in \cite{RadCorrResum}
    \item {\tt Lndecay}: from Eq.(13) in \cite{RadCorrResum}
    \item {\tt Sndecay}: from Eq.(13) in \cite{RadCorrResum}
    \item {\tt NLLndecay}:  from Eq.(14) in \cite{RadCorrResum}
\end{itemize}

In {\tt PRyM\_thermo.py}, line 116, we note the Pauli blocking correction factors for relativistic Fermions \cite{thermo-tables}
\begin{itemize}
    \item {\tt fannFD}: annihilation
    \item {\tt fscatFD}: scattering
\end{itemize}
Finally, there would be a possible dependency in the data tables from \cite{thermo-tables} used in\break {\tt PRyM\_thermo.py}, which contain the effect of finite electron mass in scattering and annihilation matrix elements and some QED plasma corrections. 

We emphasize that not all these variables have actually been changed in the code. In fact, all the mentioned variables and tables in {\tt PRyM\_eval\_nTOp.py} and {\tt PRyM\_thermo.py} have been left unperturbed. Conversely, in {\tt PRyM\_init.py} all except for {\tt gA}, {\tt radproton}, and {\tt Vud} were changed to incorporate the perturbative corrections. The unchanged values, which in particular take part in thermal corrections, have been left unchanged for two reasons. First, they have no analytic expressions  which could be subject to the perturbative treatment, neither is there experimental data from which the impact of variations could be inferred, even approximately. Second, there is no compelling reason to expect that their impact on the final observables will be significant, by comparison to the impact of the quantities which are being changed.

To implement these variations in the code, dimensionless variation functions $f$ were used for each of the parameters to be varied, which could also be called sensitivity  coefficients. Consider some parameter $Q$ for which we have the $\frac{\Delta Q}{Q}$ expression such as those in section \ref{section2}
\be
\frac{\Delta Q}{Q} = f_Q\left(R,S\right)\frac{\Delta\alpha}{\alpha}\,;\label{eq:dx}
\ee
then, with $Q_0$ being the standard value of $Q$,
\be
Q=Q_0\left(1+f_Q\frac{\Delta\alpha}{\alpha}\right)\,.\label{eq:x}
\ee
Applying this method to all the above-mentioned parameters, we have:
\begin{itemize}
    \item $f_{\alpha_{EM}}=1$
    \item $f_{m_e}=\frac{1}{2}(1+S)$, cf. Eq. (\ref{eq:dme})
    \item $f_{m_p}=0.2(1+S)+0.8R$, cf. Eq. (\ref{eq:dmp}) 
    \item $f_{Q_N}=0.1+0.7S-0.6R$, cf. Eq. (\ref{eq:dQN}) 
    \item $f_{m_n}=f_{Q_N}+\frac{m_p}{m_n}(f_{m_p}-f_{Q_N})$, cf. Eq. (\ref{eq:dmn-dmp-dQN}) 
    \item $f_{G_N}=2f_{m_p}$, cf. Eq. (\ref{eq:dGN}) 
    \item $f_{\tau_n}=-0.2-2S+3.8R$, cf. Eq. (\ref{eq:dtaun}) 
    \item $f_{G_F}=-\frac{1}{2}f_{\tau_n}$, cf. Eq. (\ref{eq:dGF-dtaun}) from \cite{Dent}
    \item $f_{G_F}=-S$, cf. Eq. (\ref{eq:dGF-dalpha}) from Coc \cite{Coc}
    \item $f_{g_p}=0.1R-0.04\left(1+S\right)$, cf. Eq. (\ref{eq:dgp for kappas})
    \item $f_{g_p}=0.12R-0.05\left(1+S\right)$, cf. Eq. (\ref{eq:dgn for kappas})
    \item $f_{m_W}=\frac{1}{2}\left(1+S\right)$, cf. Eq. (\ref{eq:dmW}) 
    \item $f_{m_Z}=\frac{1}{2}\left(2+S\right)$, cf. Eq. (\ref{eq:dmZ}) 
\end{itemize}

And for the anomalous magnetic moments, with
\bq
g_{p_0}&=&2(1+\kappa_{p_0}),\label{eq:gp0}\\
g_{n_0}&=&2\kappa_{n_0},\label{eq:gn0}
\eq
\bq
\kappa_p&=&(1+\kappa_{p_0})(1+f_{g_p}\frac{\Delta\alpha}{\alpha})-1,\label{eq:kappa_p}\\
\kappa_n&=&\kappa_{n_0}(1+f_{g_n}\frac{\Delta\alpha}{\alpha})\,.\label{eq:kappa_n}
\eq


\begin{acknowledgments}

Useful discussions with David Hilditch at various points during the development of this work are gratefully acknowledged.

This work was financed by Portuguese funds through FCT (Funda\c c\~ao para a Ci\^encia e a Tecnologia) in the framework of the project 2022.04048.PTDC (Phi in the Sky, DOI 10.54499/2022.04048.PTDC). CJM also acknowledges FCT and POCH/FSE (EC) support through Investigador FCT Contract 2021.01214.CEECIND/CP1658/CT0001 (DOI 10.54499/2021.01214.CEECIND/CP1658/CT0001). This work was partially supported by FCT Portugal through grant No. UID/PRR/00099/2025 and grant No. UID/00099/2025. This work benefited from the use of CENTRA’s cluster {\it Baltasar} and CAUP’s cluster {\it Supernova}.
\end{acknowledgments}
 
\bibliography{article}

@article{Burns,
    author = "Burns, Anne-Katherine and Tait, Tim M. P. and Valli, Mauro",
    title = "{PRyMordial: the first three minutes, within and beyond the standard model}",
    eprint = "2307.07061",
    archivePrefix = "arXiv",
    primaryClass = "hep-ph",
    reportNumber = "UCI-HEP-TR-2023-07, YITP-SB-2023-16",
    doi = "10.1140/epjc/s10052-024-12442-0",
    journal = "Eur. Phys. J. C",
    volume = "84",
    number = "1",
    pages = "86",
    year = "2024"
}

@article{Martins1,
    author = "Clara, M. T. and Martins, C. J. A. P.",
    title = "{Primordial nucleosynthesis with varying fundamental constants: Improved constraints and a possible solution to the Lithium problem}",
    eprint = "2001.01787",
    archivePrefix = "arXiv",
    primaryClass = "astro-ph.CO",
    doi = "10.1051/0004-6361/201937211",
    journal = "Astron. Astrophys.",
    volume = "633",
    pages = "L11",
    year = "2020"
}

@article{Coc,
    author = "Coc, Alain and Nunes, Nelson J. and Olive, Keith A. and Uzan, Jean-Philippe and Vangioni, Elisabeth",
    title = "{Coupled Variations of Fundamental Couplings and Primordial Nucleosynthesis}",
    eprint = "astro-ph/0610733",
    archivePrefix = "arXiv",
    reportNumber = "UMN-TH-2526-06, FTPI-MINN-06-36",
    doi = "10.1103/PhysRevD.76.023511",
    journal = "Phys. Rev. D",
    volume = "76",
    pages = "023511",
    year = "2007"
}

@article{Dent,
    author = "Dent, Thomas and Stern, Steffen and Wetterich, Christof",
    title = "{Primordial nucleosynthesis as a probe of fundamental physics parameters}",
    eprint = "0705.0696",
    archivePrefix = "arXiv",
    primaryClass = "astro-ph",
    reportNumber = "HD-THEP-07-10",
    doi = "10.1103/PhysRevD.76.063513",
    journal = "Phys. Rev. D",
    volume = "76",
    pages = "063513",
    year = "2007"
}

@article{WhiteDwarfs,
    author = "Magano, D. M. N. and Vilas Boas, J. M. A. and Martins, C. J. A. P.",
    title = "{Current and Future White Dwarf Mass-radius Constraints on Varying Fundamental Couplings and Unification Scenarios}",
    eprint = "1710.05828",
    archivePrefix = "arXiv",
    primaryClass = "astro-ph.CO",
    doi = "10.1103/PhysRevD.96.083012",
    journal = "Phys. Rev. D",
    volume = "96",
    number = "8",
    pages = "083012",
    year = "2017"
}

@article{kappas,
    author = "Ferreira, M. C. and Martins, C. J. A. P.",
    title = "{Further consistency tests of the stability of fundamental couplings}",
    eprint = "1506.03550",
    archivePrefix = "arXiv",
    primaryClass = "astro-ph.CO",
    doi = "10.1103/PhysRevD.91.124032",
    journal = "Phys. Rev. D",
    volume = "91",
    pages = "124032",
    year = "2015"
}

@MastersThesis{Thesis,
    author     =     {Dreyer, I. M.},
    title     =     {{Probing Unification Scenarios with Big Bang Nucleosynthesis}},
    school     =     {IST, University of Lisbon},
    year     =     {2025},
}

@article{Vud,
    author = "Bona, Marcella and others",
    collaboration = "UTfit",
    title = "{New UTfit Analysis of the Unitarity Triangle in the Cabibbo-Kobayashi-Maskawa scheme}",
    eprint = "2212.03894",
    archivePrefix = "arXiv",
    primaryClass = "hep-ph",
    reportNumber = "YITP-SB-2022-40",
    doi = "10.1007/s12210-023-01137-5",
    journal = "Rend. Lincei Sci. Fis. Nat.",
    volume = "34",
    pages = "37--57",
    year = "2023"
}

@article{RadCorrResum,
    author = "Czarnecki, Andrzej and Marciano, William J. and Sirlin, Alberto",
    title = "{Precision measurements and CKM unitarity}",
    eprint = "hep-ph/0406324",
    archivePrefix = "arXiv",
    reportNumber = "ALBERTA-THY-12-04, BNL-HET-04-7, NYU-TH-04-06-25",
    doi = "10.1103/PhysRevD.70.093006",
    journal = "Phys. Rev. D",
    volume = "70",
    pages = "093006",
    year = "2004"
}

@article{thermo-tables,
    author = "Escudero Abenza, Miguel",
    title = "{Precision early universe thermodynamics made simple: $N_{\rm eff}$ and neutrino decoupling in the Standard Model and beyond}",
    eprint = "2001.04466",
    archivePrefix = "arXiv",
    primaryClass = "hep-ph",
    reportNumber = "KCL-2019-85",
    doi = "10.1088/1475-7516/2020/05/048",
    journal = "JCAP",
    volume = "05",
    pages = "048",
    year = "2020"
}

@article{PDG,
    author = "Navas, S. and others",
    collaboration = "Particle Data Group",
    title = "{Review of particle physics}",
    doi = "10.1103/PhysRevD.110.030001",
    journal = "Phys. Rev. D",
    volume = "110",
    number = "3",
    pages = "030001",
    year = "2024"
}

@article{Rnegativo,
    author = "Lee, Taekoon",
    title = "{A model for time-evolution of coupling constants}",
    eprint = "2310.13308",
    archivePrefix = "arXiv",
    primaryClass = "hep-ph",
    doi = "10.1016/j.physletb.2023.138424",
    journal = "Phys. Lett. B",
    volume = "849",
    pages = "138424",
    year = "2024"
}

@article{Nakashima,
    author = "Nakashima, Masahiro and Ichikawa, Kazuhide and Nagata, Ryo and Yokoyama, Jun'ichi",
    title = "{Constraining the time variation of the coupling constants from cosmic microwave background: effect of {\textbackslash}Lambda{\_}{QCD}}",
    eprint = "0910.0742",
    archivePrefix = "arXiv",
    primaryClass = "astro-ph.CO",
    reportNumber = "RESCEU-25-09",
    doi = "10.1088/1475-7516/2010/01/030",
    journal = "JCAP",
    volume = "01",
    pages = "030",
    year = "2010"
}

@article{nacre,
    author = "Xu, Yi and Takahashi, Kohji and Goriely, Stephane and Arnould, Marcel and Ohta, Masahisa and Utsunomiya, Hiroaki",
    title = "{NACRE II: an update of the NACRE compilation of charged-particle-induced thermonuclear reaction rates for nuclei with mass number $A < 16$}",
    eprint = "1310.7099",
    archivePrefix = "arXiv",
    primaryClass = "nucl-th",
    doi = "10.1016/j.nuclphysa.2013.09.007",
    journal = "Nucl. Phys. A",
    volume = "918",
    pages = "61--169",
    year = "2013"
}

@article{primatHe4,
    author = "Pitrou, Cyril and Coc, Alain and Uzan, Jean-Philippe and Vangioni, Elisabeth",
    title = "{Precision big bang nucleosynthesis with improved Helium-4 predictions}",
    eprint = "1801.08023",
    archivePrefix = "arXiv",
    primaryClass = "astro-ph.CO",
    doi = "10.1016/j.physrep.2018.04.005",
    journal = "Phys. Rept.",
    volume = "754",
    pages = "1--66",
    year = "2018"
}

@techreport{Kawano1,
    author = "Kawano, Lawrence",
    title = "{Let's go: Early universe. 2. Primordial nucleosynthesis: The Computer way}",
    institution = "Caltech, Kellogg Lab",
    reportNumber = "FERMILAB-PUB-92-004-A",
    month = "1",
    year = "1992"
}

@article{Kawano2,
    author = "Smith, Michael S. and Kawano, Lawrence H. and Malaney, Robert A.",
    title = "{Experimental, computational, and observational analysis of primordial nucleosynthesis}",
    reportNumber = "OAP-716",
    doi = "10.1086/191763",
    journal = "Astrophys. J. Suppl.",
    volume = "85",
    pages = "219--247",
    year = "1993"
}

@article{Pisanti,
    author = "Pisanti, O. and Cirillo, A. and Esposito, S. and Iocco, F. and Mangano, G. and Miele, G. and Serpico, P. D.",
    title = "{PArthENoPE: Public Algorithm Evaluating the Nucleosynthesis of Primordial Elements}",
    eprint = "0705.0290",
    archivePrefix = "arXiv",
    primaryClass = "astro-ph",
    reportNumber = "DSF-13-07, FERMILAB-PUB-07-079-A, SLAC-PUB-12488",
    doi = "10.1016/j.cpc.2008.02.015",
    journal = "Comput. Phys. Commun.",
    volume = "178",
    pages = "956--971",
    year = "2008"
}

@article{Sarkar:1995dd,
    author = "Sarkar, Subir",
    title = "{Big bang nucleosynthesis and physics beyond the standard model}",
    eprint = "hep-ph/9602260",
    archivePrefix = "arXiv",
    reportNumber = "OUTP-95-16-P",
    doi = "10.1088/0034-4885/59/12/001",
    journal = "Rept. Prog. Phys.",
    volume = "59",
    pages = "1493--1610",
    year = "1996"
}

@article{Steigman:2007xt,
    author = "Steigman, Gary",
    title = "{Primordial Nucleosynthesis in the Precision Cosmology Era}",
    eprint = "0712.1100",
    archivePrefix = "arXiv",
    primaryClass = "astro-ph",
    doi = "10.1146/annurev.nucl.56.080805.140437",
    journal = "Ann. Rev. Nucl. Part. Sci.",
    volume = "57",
    pages = "463--491",
    year = "2007"
}

@article{Iocco:2008va,
    author = "Iocco, Fabio and Mangano, Gianpiero and Miele, Gennaro and Pisanti, Ofelia and Serpico, Pasquale D.",
    title = "{Primordial Nucleosynthesis: from precision cosmology to fundamental physics}",
    eprint = "0809.0631",
    archivePrefix = "arXiv",
    primaryClass = "astro-ph",
    reportNumber = "DSF-20-2008, FERMILAB-PUB-08-216-A, IFIC-08-37",
    doi = "10.1016/j.physrep.2009.02.002",
    journal = "Phys. Rept.",
    volume = "472",
    pages = "1--76",
    year = "2009"
}

@article{Lithium,
    author = "Fields, Brian D.",
    title = "{The primordial lithium problem}",
    eprint = "1203.3551",
    archivePrefix = "arXiv",
    primaryClass = "astro-ph.CO",
    doi = "10.1146/annurev-nucl-102010-130445",
    journal = "Ann. Rev. Nucl. Part. Sci.",
    volume = "61",
    pages = "47--68",
    year = "2011"
}

@article{Martins3,
    author = "Deal, M. and Martins, C. J. A. P.",
    title = "{Primordial nucleosynthesis with varying fundamental constants - Solutions to the lithium problem and the deuterium discrepancy}",
    eprint = "2106.13989",
    archivePrefix = "arXiv",
    primaryClass = "astro-ph.CO",
    doi = "10.1051/0004-6361/202140725",
    journal = "Astron. Astrophys.",
    volume = "653",
    pages = "A48",
    year = "2021"
}

@article{Martins2,
    author = "Martins, C. J. A. P.",
    title = "{Primordial nucleosynthesis with varying fundamental constants: Degeneracies with cosmological parameters}",
    eprint = "2012.10505",
    archivePrefix = "arXiv",
    primaryClass = "astro-ph.CO",
    doi = "10.1051/0004-6361/202039605",
    journal = "Astron. Astrophys.",
    volume = "646",
    pages = "A47",
    year = "2021"
}

@article{Luo,
    author = "Luo, Feng and Olive, Keith A. and Uzan, Jean-Philippe",
    title = "{Gyromagnetic Factors and Atomic Clock Constraints on the Variation of Fundamental Constants}",
    eprint = "1107.4154",
    archivePrefix = "arXiv",
    primaryClass = "hep-ph",
    reportNumber = "UMN-TH-3006-11, FTPI-MINN-11-16",
    doi = "10.1103/PhysRevD.84.096004",
    journal = "Phys. Rev. D",
    volume = "84",
    pages = "096004",
    year = "2011"
}

@article{Cooke,
    author = "Cooke, Ryan",
    title = "{Big Bang Nucleosynthesis}",
    eprint = "2409.06015",
    archivePrefix = "arXiv",
    primaryClass = "astro-ph.CO",
    doi = "10.1016/B978-0-443-21439-4.00046-8",
    journal = "Encyclopedia of Astrophysics",
    volume = "5",
    number = "2026",
    pages = "159-183",
    month = "8",
    year = "2025"
}

@article{primatD,
    author = "Pitrou, Cyril and Coc, Alain and Uzan, Jean-Philippe and Vangioni, Elisabeth",
    title = "{A new tension in the cosmological model from primordial deuterium?}",
    eprint = "2011.11320",
    archivePrefix = "arXiv",
    primaryClass = "astro-ph.CO",
    doi = "10.1093/mnras/stab135",
    journal = "Mon. Not. Roy. Astron. Soc.",
    volume = "502",
    number = "2",
    pages = "2474--2481",
    year = "2021"
}

@article{ANDES,
    author = "Martins, C. J. A. P. and others",
    collaboration = "ANDES",
    title = "{Cosmology and fundamental physics with the ELT-ANDES spectrograph}",
    eprint = "2311.16274",
    archivePrefix = "arXiv",
    primaryClass = "astro-ph.CO",
    doi = "10.1007/s10686-024-09928-w",
    journal = "Exper. Astron.",
    volume = "57",
    number = "1",
    pages = "5",
    year = "2024"
}

@article{other_RS_compact,
    author = "Kolonia, E. -A. and Martins, C. J. A. P. and Gourgouliatos, Konstantinos N.",
    title = "{Probing fundamental physics using compact astrophysical objects}",
    eprint = "2503.11447",
    archivePrefix = "arXiv",
    primaryClass = "hep-ph",
    doi = "10.1103/PhysRevD.111.083017",
    journal = "Phys. Rev. D",
    volume = "111",
    number = "8",
    pages = "083017",
    year = "2025"
}

@article{other_RS_clocks,
    author = "Martins, C. J. A. P.",
    title = "{Varying fundamental constants cosmography}",
    eprint = "2508.18458",
    archivePrefix = "arXiv",
    primaryClass = "astro-ph.CO",
    doi = "10.1016/j.dark.2025.102047",
    journal = "Phys. Dark Univ.",
    volume = "49",
    pages = "102047",
    year = "2025"
}

@article{Campbell,
    author = "Campbell, Bruce A. and Olive, Keith A.",
    title = "{Nucleosynthesis and the time dependence of fundamental couplings}",
    eprint = "hep-ph/9411272",
    archivePrefix = "arXiv",
    reportNumber = "UMN-TH-1318-94",
    doi = "10.1016/0370-2693(94)01652-S",
    journal = "Phys. Lett. B",
    volume = "345",
    pages = "429--434",
    year = "1995"
}

@article{Murphy,
    author = "Murphy, Michael T. and others",
    title = "{Fundamental physics with ESPRESSO: Precise limit on variations in the fine-structure constant towards the bright quasar HE 0515{\ensuremath{-}}4414}",
    eprint = "2112.05819",
    archivePrefix = "arXiv",
    primaryClass = "astro-ph.CO",
    doi = "10.1051/0004-6361/202142257",
    journal = "Astron. Astrophys.",
    volume = "658",
    pages = "A123",
    year = "2022"
}

\end{document}